\documentclass{IEEEtaes}

\usepackage{color,array,amsthm}
\usepackage{graphicx}
\usepackage{amsmath}
\usepackage{amssymb}
\usepackage{float}
\usepackage{bbold}
\usepackage{subcaption}
\usepackage{eucal}
\usepackage{mathtools} 
\usepackage{cite}
\usepackage{hyperref}
\newtheorem{theo}{Theorem}

\newtheorem{definition}{Definition}
\newtheorem{rmk}{Remark}
\newtheorem*{problem}{Problem}
\newtheorem{example}{Example}

\begin{document}

\title{Consensus-based formation of a swarm of quadrotors interacting over ring digraphs}

\author{Sahaya Aarti Dennisselvan}
\member{Student Member, IEEE}
\affil{Indian Institute of Technology Bombay, India} 

\author{Shashi Ranjan Kumar}
\member{Senior Member, IEEE}
\affil{Indian Institute of Technology Bombay, India}

\author{Dwaipayan Mukherjee}
\member{Member, IEEE}
\affil{Indian Institute of Technology Bombay, India}

 \receiveddate{Manuscript received XXXXX 00, 0000; revised XXXXX 00, 0000; accepted XXXXX 00, 0000.\\
``This work was supported in part by an Anusandhan National Research Foundation (ANRF) grant with code CRG/2023/002280.'' }
\accepteddate{XXXXX XX XXXX}
 \publisheddate{XXXXX XX XXXX}

\corresp{ {\itshape (Corresponding author: S. A. Dennisselvan)}. }

\authoraddress{ Sahaya Aarti Dennisselvan is with the Department of Electrical Engineering, Indian Institute of Technology Bombay, Powai, Mumbai- 400076, India
(e-mail: \href{sahayaaarti@ee.iitb.ac.in}{sahayaaarti@ee.iitb.ac.in}). Shashi Ranjan Kumar is with the Department of Aerospace Engineering, Indian Institute of Technology Bombay, Powai, Mumbai- 400076, India (e-mail: \href{srk@aero.iitb.ac.in}{srk@aero.iitb.ac.in}). Dwaipayan Mukherjee is with the Department of Electrical Engineering, Indian Institute of Technology Bombay, Powai, Mumbai- 400076, India
(e-mail: \href{dm@ee.iitb.ac.in}{dm@ee.iitb.ac.in}).}

 \editor{Mentions of supplemental materials and animal/human rights statements can be included here.}
\supplementary{Color versions of one or more of the figures in this article are available online at \href{http://ieeexplore.ieee.org}{http://ieeexplore.ieee.org}.}

%\markboth{AUTHOR ET AL.}{SHORT ARTICLE TITLE}
\maketitle

\begin{abstract}
This work proposes a cooperative strategy for a group of quadrotors interacting over ring digraphs with macro-vertices of size two. Consensus for a group of general double integrators has been initially investigated, and it has been proved that through a suitable choice of a single controller parameter, consensus and stability of the resulting networked dynamical system can be ensured. This further opens up the possibility of achieving a desired formation and to move a swarm of quadrotors, interacting over ring digraphs, at a desired flight velocity, using a single controller gain. An analysis of achievable velocities is performed. Examples have been provided to offer deeper insights into the obtained analytical results. Simulation studies clearly demonstrate that a desired formation is achieved, starting from arbitrary initial positions, while also ensuring convergence to a final desired flight velocity.
\end{abstract}

\begin{IEEEkeywords}
Consensus, formation, quadrotors, digraphs  
\end{IEEEkeywords}
\newpage
\section{INTRODUCTION}
Unmanned aerial vehicles (UAVs) have garnered attention within the control community due to their simplicity, compactness, and ability to perform coordinated tasks, making them applicable in surveillance \cite{borkar2020reconfigurable}, coverage control \cite{funada2023distributed}, and defense applications, among other areas. A fundamental coordination problem in swarms is the formation control problem, where agents arrange themselves in a predetermined shape. Centralized strategies are typically impractical for real-world settings, motivating the adoption of distributed and decentralized algorithms.

Recent advances in consensus-based formation control for quadrotor UAVs have addressed obstacle avoidance \cite{wu2020new}, state constraints \cite{wang2024controllers, restrepo2022robust, cui2023fixed}, disturbances \cite{li2021appointed}, faults \cite{zhou2021fixed}, and actuator failures \cite{wang2024event}. Other approaches have also been explored, although most rely on positional and/or attitude consensus \cite{gudeta2023consensus,lin2023aggressive,jin2024adaptive,xu2023adaptive}. Leader–follower structures using directed networks have been widely used; for example, \cite{jin2024adaptive} analyzed their performance under heterogeneous control parameters. A linear–quadratic optimal consensus algorithm for formation control and trajectory tracking was proposed in \cite{du2017distributed}. Additional leader–follower methods include heading-based feedback control \cite{wang2020UAVS} and local observer–based LQR control \cite{zaidi2021distributed}. A distributed, robust, and safety-critical strategy for leader–follower formation tracking was introduced in \cite{xia2025distributed}.

A safe, reliable, distributed fault-tolerant scheme addressing external disturbances, failures, and collision avoidance was proposed in \cite{hu2025safe}. Several other effective fault-tolerant schemes for formation tracking have also been reported \cite{xu2024fault, yang2024predefined, chen2023human}. A global optimal consensus problem involving bounded inputs and individual time-varying quadratic objectives for each UAV was formulated in \cite{yang2024global}. A common feature among most of these works is their reliance on leader–follower structures, which may be impractical due to communication constraints and vulnerability to leader failure. In contrast, leaderless topologies involving directed cycles, as exemplified by the cyclic pursuit paradigm, provide a fully distributed alternative and have gained substantial interest.

Cyclic pursuit, one of the classical consensus algorithms which was initially inspired by the `$n$-bug' problem, has widely attracted the attention of many researchers over the years \cite{marshall2004formations,sinha2006generalization, mukherjee2013reachability, kumar2019cooperative}. Single as well as double integrator models were used within the cyclic pursuit paradigm. In general, the necessary and sufficient condition for second-order consensus over a directed network was established in \cite{ren2007distributed,zhu2009general, yu2010some}, based on explicit knowledge of the Laplacian spectrum. However, for general weighted digraphs, evaluation of such a spectrum is not always tractable in a distributed manner. For the specific case of cyclic pursuit, an analysis of the reachability set for double integrators was carried out in \cite{mukherjee2013reachability}, based on edge weights/gains, which are local variables for an agent. A generalization of cyclic pursuit to consider double integrators was presented using negative real-valued edge weights, and was later extended to handle heterogeneous inputs and communication delays \cite{ de2023rendezvous}. These studies, however, incorporated absolute velocity of agents in their control laws.  

Hierarchical cyclic pursuit, with heterogeneous gains, was introduced in \cite{mukherjee2016generalized}, and its stability was examined. The authors also presented conditions to ensure global reachability of agents in cyclic pursuit. However, the agents were modeled as single integrators. Ring digraphs involve another special class of hierarchical cycle structures which were introduced by \cite{parsegov2023hierarchical}. The unweighted Laplacian's spectra, for such topologies, was analyzed in \cite{parsegov2023hierarchical}. Subsequently, the effect of antagonistic interactions with real-valued and non-identical edge weights over ring digraphs was analyzed in \cite{11108093}, with agents being modeled as single integrators. However, interactions of agents exhibiting more complicated dynamics, over such ring digraphs, remain largely unexplored.

The dynamics of several vehicles, such as quadrotors, can be realistically approximated as double integrators, thereby motivating the analysis of double integrators interacting over ring digraphs. In view of the open problems discussed above, we propose a decentralized control strategy to move a swarm of quadrotors, interacting over a ring digraph, with a common desired velocity, thereby attaining a geometric shape. Since the positional dynamics of a quadrotor can be approximately modeled as a double integrator, we first analyze the stability of a consensus protocol on double integrators over a weighted ring digraph. We also demonstrate that by adjusting a
single control gain, the stability may be ensured and direction of the final
flight can be chosen. We further obtain the set of achievable velocities, given a set of initial velocities, by choosing a single control gain. Several examples, along with illustrative figures, are presented to validate the theoretical results. Simulation studies have been performed considering the nonlinear model of quadrotors, to demonstrate that the group achieves the desired formation and flight velocity with the proposed choice of control input.

\par{Notation:} We define the set of real numbers as $\mathbb{R}$ and $\iota=\sqrt{-1}$ denotes the imaginary unit. A column vector $\mathbb{1}_p$ consisting of ones in $\mathbb{R}^p$ and $\mathbb{0}_p$ denotes a vector of zeros in $\mathbb{R}^p$.  The terms $ I_{ N\times N}$ and $\mathbb{0}_{ N\times N}$ refer to identity and zero matrices of size $ \ N\times N$, while $\otimes$ denotes the Kronecker product. For a matrix $A\in\mathbb{C}^{p\times q}$, the matrix $A^{*}\in\mathbb{C}^{p\times q}$ is obtained by entry-wise conjugation of the matrix $A$.

\section{PRELIMINARIES}
The collective behavior of a swarm of quadrotors is represented using graphs, wherein each quadrotor is defined as a \textit{node} or \textit{vertex} and their interactions are captured through \textit{edges}. A directed graph (digraph) $\mathcal{G} = (\mathcal{V}, \mathcal{E})$ consists of a finite set of vertices, $\mathcal{V} = \{ \mathfrak{v}_{1}, \mathfrak{v}_{2}, \ldots, \mathfrak{v}_{n} \}$, and a set of directed edges, $\mathcal{E} \subseteq \mathcal{V} \times \mathcal{V}$. In a weighted digraph, each edge is associated with a corresponding real-valued weight. Presence of an edge $(\mathfrak{v}_i, \mathfrak{v}_j) \in \mathcal{E}$ in a graph indicates agent $\mathfrak{v}_i$ `chases' agent $\mathfrak{v}_j$, implying that agent $\mathfrak{v}_i$ has relative state information with respect to agent $\mathfrak{v}_j$.
A directed path is a sequence of ordered edges of the form $\{(\mathfrak{v}_{i_1}, \mathfrak{v}_{i_{2}}),\ldots,(\mathfrak{v}_{i_p}, \mathfrak{v}_{i_{p+1}}),(\mathfrak{v}_{i_{p+1}}, \mathfrak{v}_{i_{p+2}}),\ldots, (\mathfrak{v}_{i_j}, \mathfrak{v}_{i_{j+1}})\}$. A directed cycle is a directed path whose starting and terminal vertices are identical, i.e., $\mathfrak{v}_{i_1}=\mathfrak{v}_{i_{j+1}}$. A directed tree with exactly one vertex, called the \textit{root}, having out-degree zero, while all other vertices have out-degree one, is termed a rooted tree/rooted in-branching. A digraph is said to possess a spanning tree if there exists a vertex to which a directed path exists from every other vertex.  With this vertex now being the root for such a spanning tree within the graph, the digraph thus possesses a rooted spanning tree.

An (out) adjacency matrix $A_N \in \mathbb{R}^{N \times N}$ of a digraph is denoted by $[A_N]_{ij}= a_{ij}$ where $a_{ij}$ equals the number of arcs of the form $(\mathfrak{v}_i,\mathfrak{v}_j) \in \mathcal{E}$. The graph (out) Laplacian $\mathcal{L} \in \mathbb{R}^{N \times N}$ associated with a digraph is given by its entries $[\mathcal{L}]_{ii}\coloneqq \ell_{ii}=\sum_{j \neq i} a_{ij}$ and $[\mathcal{L}]_{ij}\coloneqq \ell_{ij}= - a_{ij}$ for $i \neq j$. Further details about the graph theory may be found in \cite{godsil2001algebraic}. In this paper, we shall consider the network topology proposed in \cite{parsegov2023hierarchical}. Some relevant definitions pertaining to the same are presented below:
\begin{definition}(\cite{parsegov2023hierarchical})
    A linear macro-vertex $\mathcal{G}_{n}^{i}=\left( \mathcal{V}^{i},\mathcal{E}^{i} \right)$ of a digraph $\mathcal{G}_{N}=(\mathcal{V},\mathcal{E})$ is a sub-digraph of $\mathcal{G}_{N}$ with $\mathcal{V}^{i}=\left\{ \mathfrak{v}_{1}^{i},\mathfrak{v}_{2}^{i},\dots,\mathfrak{v}_{n}^{i} \right\} (n \ge 1)$ obtained from the directed path $\mathfrak{v}_{n}^{i}\to \mathfrak{v}_{n-1}^{i}\cdots \to \mathfrak{v}_{1}^{i}$ (main direction; no arcs when $n =1$) by adding the reverse path $\mathfrak{v}_{1}^{i}\to \mathfrak{v}_{2}^{i}\cdots \to \mathfrak{v}_{n}^{i}$ from which any subset of arcs is dropped.
\end{definition}
\begin{definition}(\cite{parsegov2023hierarchical})
    A ring digraph denoted by $\mathcal{G}_{m,n}=(\mathcal{V},\mathcal{E})$ is a digraph such that $\mathcal{V}=
\bigcup_{i=1}^{m}\mathcal{V}^{i}$, $\mathcal{V}^{i}=\left\{ \mathfrak{v}_{n(i-1)+1},\dots,\mathfrak{v}_{ni} \right\}$, $\mathcal{E} =\left( \bigcup_{i=1}^{m}\mathcal{E}^{i} \right) \cup \left\{ e_{1}, \dots, e_{m}\right\}$, $\left( \mathcal{V}^{i},\mathcal{E}^{i} \right)=\mathcal{G}_{n}^{i}$  are identical linear macro-vertices on $n \ge 1$ nodes, and the arcs $e_{i}=\left( \mathfrak{v}_{ni+1},\mathfrak{v}_{ni} \right) \left( i\in \left\{ 1,\dots ,m-1 \right\} \right)$ and $e_{m}=\left( \mathfrak{v}_1,\mathfrak{v}_{nm} \right)$ link the first node of each macro-vertex with the $n$-th node of the previous one (which is the same macro-vertex when $m = 1$).
\end{definition}

The necessary and sufficient condition for the consensus among double integrators is stated below.

\begin{theo}(\cite{yu2010some}) \label{yu_bound}
Consensus among double integrators is achieved if and only if the network has a directed spanning tree and
\begin{align}
    \frac{\beta^{2}}{\alpha} >   \max_{2\le i\le N}\frac{\mathfrak{Im}^{2} (\mu_i)}{\mathfrak{Re}(\mu_i) \left[\mathfrak{Re}^2(\mu_i) +\mathfrak{Im}^{2} (\mu_i)  \right]},
\end{align}
where $\mu_i$ are the non-zero eigenvalues of the Laplacian, while $\mathfrak{Re}(\cdot)$ and $\mathfrak{Im}(\cdot)$ represent the real and imaginary part of any complex number, respectively.  
\end{theo}

Another existing result that will aid our analyses of the Laplacian spectra for a ring digraph is stated below.
\begin{theo}(\cite{davis1979circulant})\label{circulant_theo}
    A matrix $\mathcal{C} \in \mathbb{C}^{2m \times 2m}$ is said to be block circulant if and only if it is of the form 
    \begin{align}
        \mathcal{C} =\bar{F}  \text{diag}(M_1,M_2,\dots, M_m)  F,\;\; F=(F_{m} \otimes  F_{2})
    \end{align}
    where $M_{k}$ are arbitrary matrices in $\mathbb{C}^{2 \times 2}$. Also, $F_m$ and $F_2$ are Fourier matrices in $\mathbb{C}^{m \times m}$ and $\mathbb{C}^{2 \times 2}$ respectively, while $\bar{F}\coloneqq (F^T)^*=F^{-1}$, since $F$ is unitary matrix. 
\end{theo}
%%%%%%%%%%%%%%%%%%%%%%
\section{PROBLEM FORMULATION}
\begin{figure}
    \centering
    \includegraphics[width=0.75\linewidth]{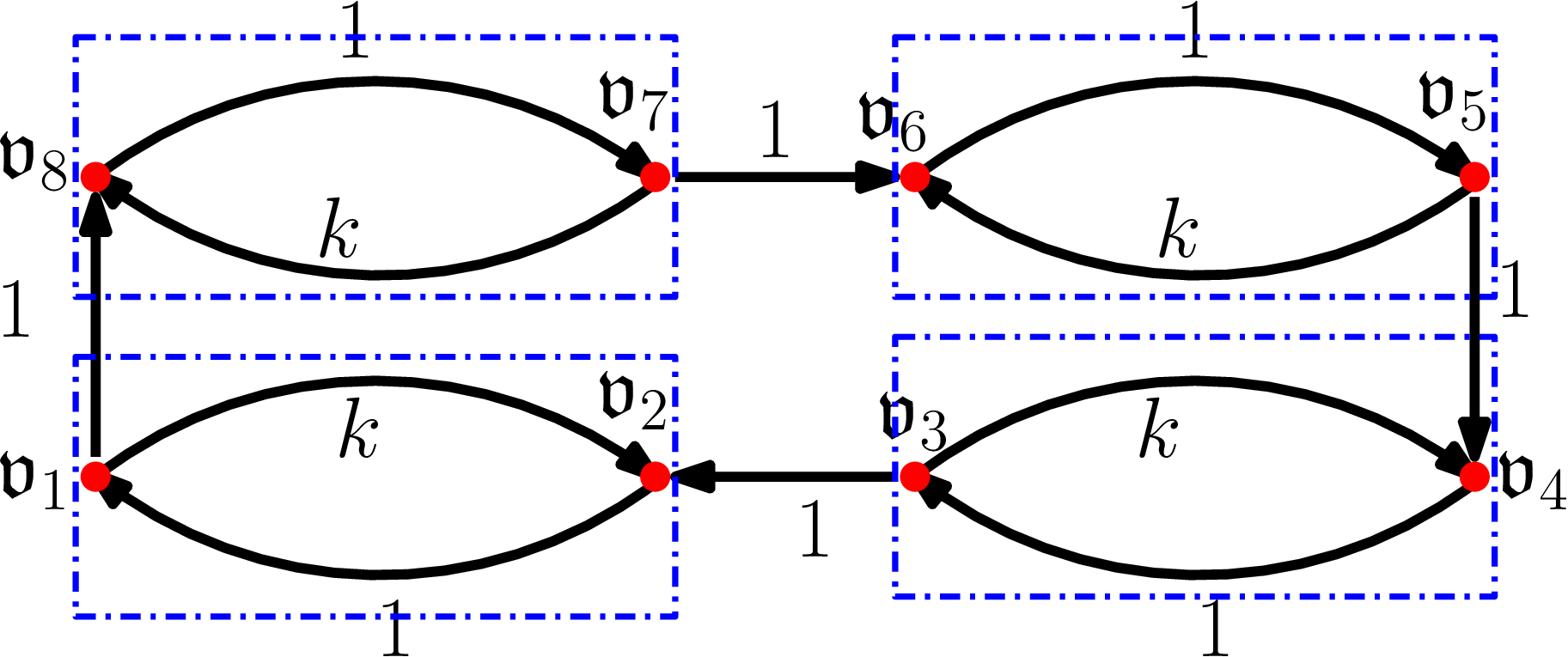}
    \caption{Ring digraph with $m=4$.}
    \label{example m=4}
\end{figure}

Consider a group of $N$ quadrotors interacting in a decentralized setup using the Euler angle-based model. Suppose the quadrotors interact over ring digraphs with $N=2m$, where $m$ is the number of macro-vertices. A ring digraph with $m=4$ is shown in Fig. \ref{example m=4}.
The dynamics of each quadrotor are represented in the following form:
\begin{align}\label{position dynamics}
    \text{Position dynamics:} \begin{cases}
    \dot{p}_i&=v_i;\\
 \dot{v}_i&= \frac{T_i}{m_i} R_i e_3 - ge_3, \\
    \end{cases}
\end{align}
\begin{align} \label{attitude dynamics}
\text{Attitude dynamics:} \begin{cases}
\dot{\Theta}_i &= W_i \omega_i; \\
J_i \dot{\omega}_i &=\tau_i,
    \end{cases}
\end{align}
where $i\in \left\{ 1,2,\dots, N \right\}$, $m_i$ denotes the $i^{\text{th}}$ quadrotor's mass, $J_i \in \mathbb{R}^{3 \times 3}$ denotes its symmetric and positive definite inertia matrix, and $p_i =[x_i~ y_i~ z_i]^T \in \mathbb{R}^3$ and $v_i \in \mathbb{R}^3$ denote its position and linear velocity with respect to earth-fixed inertial frame, respectively. Let $\Theta_i\coloneqq [\phi_i~\theta_i~\psi_i]^{T}$ represent its roll, pitch, and yaw angles. Further, let $\omega_i \in \mathbb{R}^3$ denote its angular velocity, and suppose the rotation matrix $R_i \in SO(3)$ transforms the vectors from the body reference frame to an inertial frame. The matrix $W_i$ is given by $W_i \coloneqq \begin{bmatrix}
1 & \tan \theta_i \sin \phi_i & \tan \theta_i \cos \phi_i \\
 0& \cos \phi_i & -\sin \phi_i  \\
0 & \sin \phi_i / \cos \theta_i  & \cos \phi_i / \cos \theta_i
\end{bmatrix}$. 
The inputs $T_i$ and $\tau_i$ represent external thrusts and moments generated by the propellers in the body
axes. Details about a quadrotor's rigid body kinematics and dynamics may be found in \cite{quan2017introduction}. 

Quadrotors possess very low rotational inertia and can generate high torque, $\tau_i$. As a result, the time constant of the position dynamics in \eqref{position dynamics} is much larger than that of the attitude dynamics in \eqref{attitude dynamics}. This separation in timescales \cite{dong2014time} makes it reasonable to approximate the position dynamics \eqref{position dynamics} by a double integrator model given by
\begin{equation}\label{double_inte} 
        \dot{p}_i (t) =v_i (t),~
        \dot{v}_i (t)=u_i (t), 
\end{equation}
where $p_i, v_i \in \mathbb{R}^{\mathfrak{d}}$ are the position and velocity of $i^{\text{th}}$ agent in $\mathfrak{d}$-dimensional space.
We are interested in designing cooperative laws based on relative state information with respect to neighbors. We adopt the following control law:
\begin{align}
\nonumber
    u_i (t)=& - \alpha\sum_{j \in \mathcal{N}_i}^{} a_{ij} \left( p_i(t)-p_j(t) \right) \\ 
     &-\beta \sum_{j \in \mathcal{N}_i}^{} a_{ij} \left( v_i(t)-v_j(t) \right), \label{control}
\end{align}
where $a_{ij}$ is the $(i,j)^{\text{th}}$ entry of the weighted adjacency matrix, while $\alpha>0$, and $\beta >0$ are the coupling strengths. In compact form, the dynamics under this control law may be written as
\begin{equation}\label{overall eqn} 
        \dot{p} (t) =v (t),~~
        \dot{v} (t)= - \alpha(\mathcal{L} \otimes I_{\mathfrak{d}}) p (t) - \beta (\mathcal{L} \otimes I_{\mathfrak{d}}) v(t), 
\end{equation}
where $p\coloneqq \begin{bmatrix}
p_1^T & p_2^T & \cdots  & p_N^T
\end{bmatrix}^{T}$, $v\coloneqq \begin{bmatrix}
v_1^T & v_2^T & \cdots  & v_N^T
\end{bmatrix}^{T}$, and $\mathcal{L}$ denotes the (out) Laplacian of the overall directed network. Here, we consider $\mathfrak{d}=1$ throughout, and our results extend readily to higher dimensions. Define $Z(t)\coloneqq  \begin{bmatrix}
p^T(t) & v^T(t) 
\end{bmatrix}^{T}$ (with $\mathfrak{d}=1$), so that \eqref{overall eqn} turns out as follows:
\begin{align}\label{closedloop}
    \dot{Z}(t) = \tilde{\mathcal{L}} Z(t),~  \tilde{\mathcal{L}} \coloneqq\begin{bmatrix}
\mathbb{0}_{N \times N} & I_{N \times N} \\
 -\alpha\mathcal{L}&  - \beta \mathcal{L}
\end{bmatrix}.
\end{align}
Given the relationship between position dynamics and cooperative law for double integrators, as presented in \eqref{control}, our objective is to achieve consensus in the velocities of double integrators over ring digraphs, which can then be tailored to achieve some desired formation of a swarm of quadrotors.
Thus, the main problem addressed in this work is as follows:

\begin{problem}
Design a decentralized cooperative strategy with minimal control parameters such that multiple quadrotors interacting over a ring digraph form a desired shape and, in addition, move along a common desired direction.     
\end{problem} 
%%%%%%%%%%%%%%%%%%%%%%%%%%%%%%%%%%%%%%%%%%%%%%%%%%%%%%%%%%%%%
\section{MAIN RESULTS} \label{main results}
We first present the results for double integrator and then tailor it for achieving the formation of quadrotors. 
\subsection{Consensus of double integrators}
The ring digraphs in \cite{parsegov2023hierarchical} considered unweighted edges, signifying equal influence among all the neighbors. 
We introduce a control gain $k \in \mathbb{R}$ in every macro-vertex, which is then fused to form a ring digraph, as shown in Fig. \ref{example m=4}. 
The graph Laplacian, $\mathcal{L}\in\mathbb{R}^{N\times N}$, for this network is given by
\begin{equation}\label{laplacian}
    \mathcal{L}=\begin{bmatrix}
1+k & -k & 0 & 0 & \cdots  & 0 & -1 \\
 -1&1  & 0 & 0 & \cdots  & 0 & 0  \\
 0& -1 & 1+k  &-k  &  \cdots&0&0  \\
 \vdots & \ddots  & \ddots  & \ddots  & \ddots  &  \ddots &\vdots \\
 0& \cdots  & 0 & -1 & 1 & 0 & 0 &  \\
  0 &\cdots   & 0 &0& -1  & 1+k &-k  \\
  0 & \cdots  & 0 &0 & 0& -1 & 1
\end{bmatrix}.
\end{equation}
\begin{rmk}
    From the topology itself, it is apparent that the network contains a rooted spanning tree, and hence, the Laplacian considered in \eqref{laplacian} has exactly one eigenvalue at zero with $\mathbb{1}_N$ being the corresponding eigenvector.
\end{rmk}
 From \eqref{laplacian}, it is evident that the control gain $k$ plays a vital role in deciding the Laplacian spectra. As per Theorem \ref{yu_bound}, we need to ensure that the non-zero eigenvalues of $\mathcal{L}$ are in the right half of the complex plane, and obtain conditions on $\alpha$ and $\beta$ thereafter to ensure consensus of the system in \eqref{double_inte} under the control law \eqref{control}.
 Therefore, this requires us to ascertain the range of control gain, $k$, that ensures the non-zero eigenvalues of $\mathcal{L}$ have positive real parts or $-\mathcal{L}$ has all its non-zero eigenvalues in the left half of the complex plane.

The matrix $-\mathcal{L}$ has a special block circulant structure, with blocks of size $2 \times 2$, given by
\begin{align*}
   -\mathcal{L}= \begin{bmatrix}
B_1 & B_2 & \cdots  & B_m \\
B_m & B_1 & \cdots  & B_{m-1} \\
 \vdots & \vdots  & \ddots  & \vdots  \\
B_2 & B_3 & \cdots & B_1
\end{bmatrix}, 
\end{align*}
where $B_1= \begin{bmatrix}
    -(1+k) & k \\1 &-1
\end{bmatrix}$, $B_m=\begin{bmatrix}
    0 & 1 \\0 &0
\end{bmatrix}$, and $B_2$ through $B_{m-1}$ are zero matrices in $\mathbb{R}^{2\times 2}$. Any block circulant matrix, say $\mathcal{C}$, can be block diagonalized using the Fourier matrices, ${F}_{n}\coloneqq \dfrac{1}{\sqrt{n}} \begin{bmatrix}
1 & 1 & 1 &  \cdots & 1 \\
 1& \omega & \omega^{2} & \cdots  & \omega^{n-1} \\
 1& \omega^{2} & \omega^{4} & \cdots  & \omega^{2(n-1)} \\
 \vdots & \ddots   & \ddots  & \ddots  & \vdots  \\
1 & \omega^{n-1} & \omega^{2(n-1)}  & \cdots  & \omega^{(n-1)(n-1)}
\end{bmatrix},$ with $\omega=e^{-\iota({2\pi}/{n})}$, as stated in Theorem \ref{circulant_theo}.
Using Theorem \ref{circulant_theo}, the graph Laplacian may thus be block diagonalized and thereafter represented as follows:
\begin{equation}\label{eq.diag}
    \mathcal{D}= {F} (-\mathcal{L}) \bar{F};\;\; F=(F_{m} \otimes  F_{2}).
\end{equation}
The block diagonal matrix $\mathcal{D}$ has $m$ blocks each in $\mathbb{C}^{2 \times 2}$. Thus, it can be represented as $\mathcal{D}=\text{blockdiag}(\mathcal{D}_{1}, \mathcal{D}_{2}, \dots, \mathcal{D}_{m})$, where each block $\mathcal{D}_{i} \in \mathbb{C}^{2 \times 2}$.
The block diagonal matrix $\mathcal{D}$ follows a special structure, wherein the first block, $\mathcal{D}_1$, a real matrix, has two real eigenvalues, including a zero eigenvalue, irrespective of $m$, which follows from the fact that
\begin{align*} \mathcal{D}_{1}=\begin{bmatrix}
0 & -k \\
 0&-(k+2) 
\end{bmatrix}.
\end{align*}
The other blocks $\mathcal{D}_i,~ i \neq 1$ are possibly complex matrices. We note that blocks $\mathcal{D}_{2}$ through $\mathcal{D}_{m}$ are such that $\mathcal{D}_{\ell}=\mathcal{D}_{m+2-\ell}^*$ for $\ell\in\{2,3,\ldots, m\}$. Clearly, all blocks except $\mathcal{D}_{1}$ and $\mathcal{D}_{(m/2)+i}$ (when $m$ is even) appear in complex conjugate pairs, and the corresponding eigenvalues of these conjugate blocks also appear in complex conjugate pairs. For even values of $m$, $\mathcal{D}_{(m/2)+i}$ also has real entries. The eigenvalues of each block, $ \mathcal{D}_{i}$, are thus distinct but possibly complex. 
These conclusions regarding the structures and spectra of the diagonal blocks follow from an explicit diagonalization using \eqref{eq.diag}. The detailed steps are omitted for brevity.
As an illustration, it may be observed that the $\ell$-th block, $\mathcal{D}_{\ell}$, is given by
\begin{align}\label{eq.d2}
    \mathcal{D}_{\ell}=\begin{bmatrix}
-\alpha_{\ell} & -k+\alpha_{\ell} \\
- \alpha_{\ell}& -k-2+\alpha_{\ell} 
\end{bmatrix}, \; \forall \ell= \left\{2,\dots,m\right\},
\end{align}
where $\alpha_{\ell}=\left[ 1-\omega_{\ell}  \right]/2$
is complex with $\omega_{\ell}=e^{-\iota\frac{2\pi(\ell-1)}{m}}$.
\begin{theo} \label{k_theorem}
All the non-zero eigenvalues of the graph Laplacian $\mathcal{L}$ represented in \eqref{laplacian} have positive real parts if and only if $k> -2 + \sqrt{2} \cos \left(\dfrac{\pi}{m}\right)$.
\end{theo}
\begin{proof}
Ensuring all the non-zero eigenvalues of $\mathcal{L}$ have positive real parts is equivalent to ensuring $-\mathcal{L}$ has all its non-zero eigenvalues in the left half of the complex plane. 
    As a result of the block diagonalization of $-\mathcal{L}$, we have $\mathcal{D}$, whose constituent diagonal blocks are
    \begin{align}
    \mathcal{D}_{\ell}=\begin{bmatrix}
-\alpha_{\ell} & -k+\alpha_{\ell} \\
 -\alpha_{\ell}& -k-2+\alpha_{\ell} 
\end{bmatrix}, \; \forall \ell= \left\{1,\dots,m\right\},
\end{align}
where $\alpha_{\ell}=\left[ 1-\omega_{\ell} \right]/2$
is complex with $\omega_{\ell}=e^{-\iota\frac{2\pi(\ell-1)}{m}}$. The characteristic polynomial of $\mathcal{D}_{\ell}$ is given by
\begin{align}\label{Eq.d2charpoly}
    \mathcal{X}_{\mathcal{D}_{\ell}}(s)= s^{2}+(2+k)s+2\alpha_{\ell}, \;\forall \ell= \left\{1,\dots,m\right\},
\end{align}
where $\alpha_{\ell}=[1-\omega_{\ell}]/2$ is a complex number. This leads to the following characteristic equation:
\begin{align}\label{charpoly_eq}
    s^{2}+(2+k)s+ 2 \sin^{2}\left( \frac{\theta_{\ell}}{2} \right) + \iota 2\sin\left( \frac{\theta_{\ell}}{2} \right) \cos \left( \frac{\theta_{\ell}}{2} \right) =0,
\end{align}
where $\theta_{\ell}=\frac{2\pi(\ell-1)}{m}$. Equation \eqref{charpoly_eq} can be further expressed as the characteristic equation of a closed-loop system as follows: 
\begin{equation} \label{k_closed}
    1+k \underbrace{ \frac{s}{s^{2}+2s+ 2 \sin^{2}\left( \frac{\theta_{\ell}}{2} \right) +  2 \iota \sin\left( \frac{\theta_{\ell}}{2} \right) \cos \left( \frac{\theta_{\ell}}{2} \right)}}_{G_{\ell}(s)} =0.
\end{equation}
Since $k$ can take any real values, we analyze \eqref{k_closed} as two separate cases.\\
\textit{Case 1:} We first consider the case when $k >0$. We will now investigate how roots of $1+kG_{\ell}(s)$ vary with changes in $k$. Equation \eqref{k_closed} is already in  Evan's form and thus lends itself to an analysis using root locus. The poles of $G_{\ell}(s)$ are the roots of $D_{\ell}(s)=s^{2}+2s+ 2 \sin^{2}\left( \frac{\theta_{\ell}}{2} \right) + \iota 2\sin\left( \frac{\theta_{\ell}}{2} \right) \cos \left( \frac{\theta_{\ell}}{2} \right)$, which are given by
\begin{align*}
    s &= -1 \pm \sqrt{\cos(\theta_{\ell})- \iota\sin(\theta_{\ell})}\\
    &= -1 \pm \exp ^{- \iota \frac{\theta_{\ell}}{2}},~~\forall \ell =\{ 1,\dots,m\}
\end{align*}
These roots of $\mathcal{D}_{\ell}$ may not be complex conjugates to each other but are, nonetheless, located in the left half of the complex plane. The numerator of $G_{\ell}(s)$ has one root at the origin. For each such block, $\mathcal{D}_{\ell}$, of $\mathcal{D}$, there are thus two complex poles and one zero at the origin. Clearly, we need to use the root locus for polynomials with complex coefficients \cite{doria2013root} to determine the range of $k$ for stability.
Since the polynomial involves complex coefficients, the root locus need not be symmetric about the real axis, but other rules, based on application of the angle criterion, still apply.
The root locus asymptote branches out in the left half of the complex plane. All possible pole-zero maps and corresponding root loci, for different values of $\ell$, fall under the four scenarios depicted in Fig. \ref{complementary RL}.
\begin{figure}[t]
    \centering
    % Row 1
    \subfloat{%
        \includegraphics[width=0.4\linewidth]{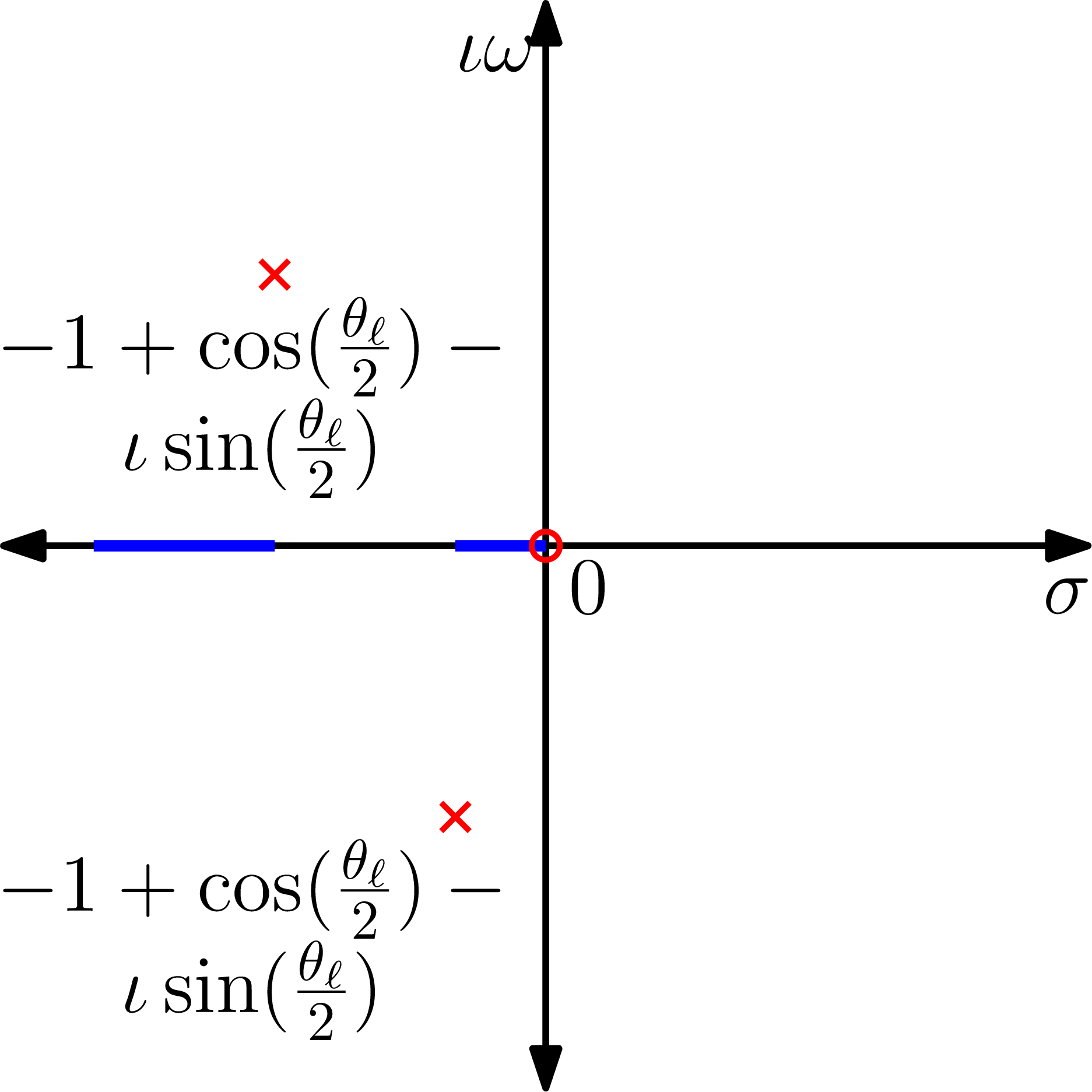}%
        \label{xy}
    }\hfill
    \subfloat{%
        \includegraphics[width=0.4\linewidth]{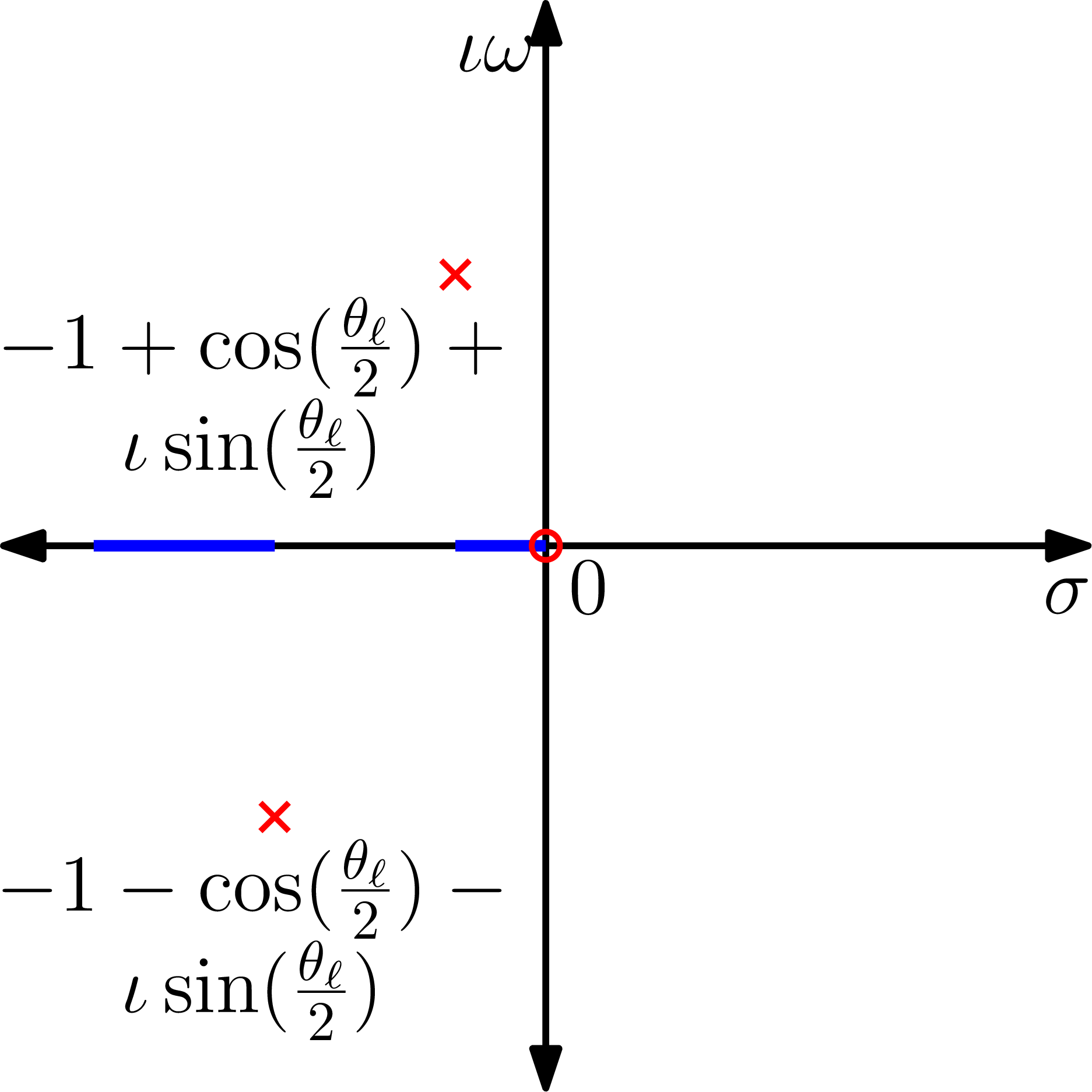}%
        \label{Time-to-go}
    }\\[1ex]
    % Row 2
    \subfloat{%
        \includegraphics[width=0.4\linewidth]{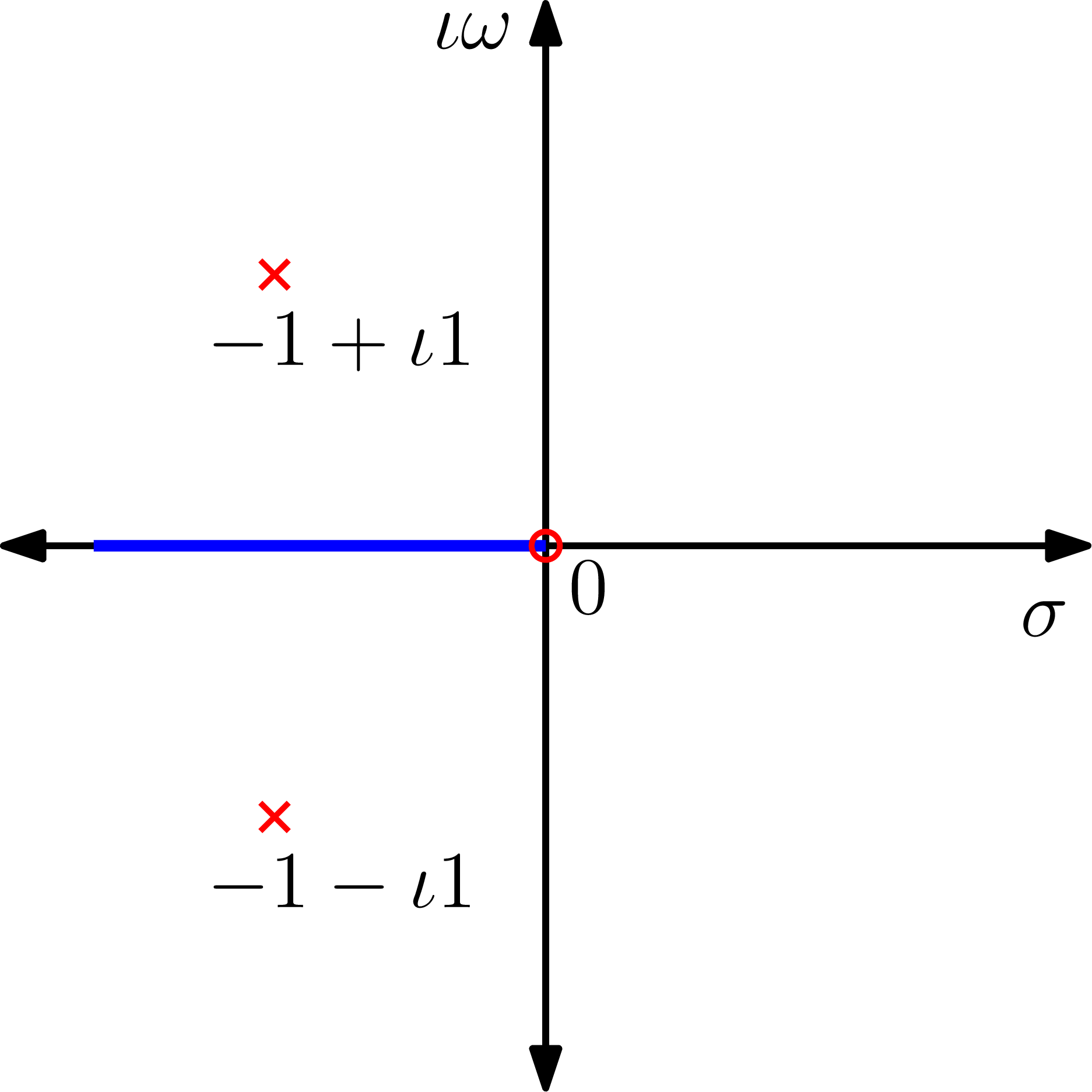}%
        \label{a_sigma}
    }\hfill
    \subfloat{%
        \includegraphics[width=0.4\linewidth]{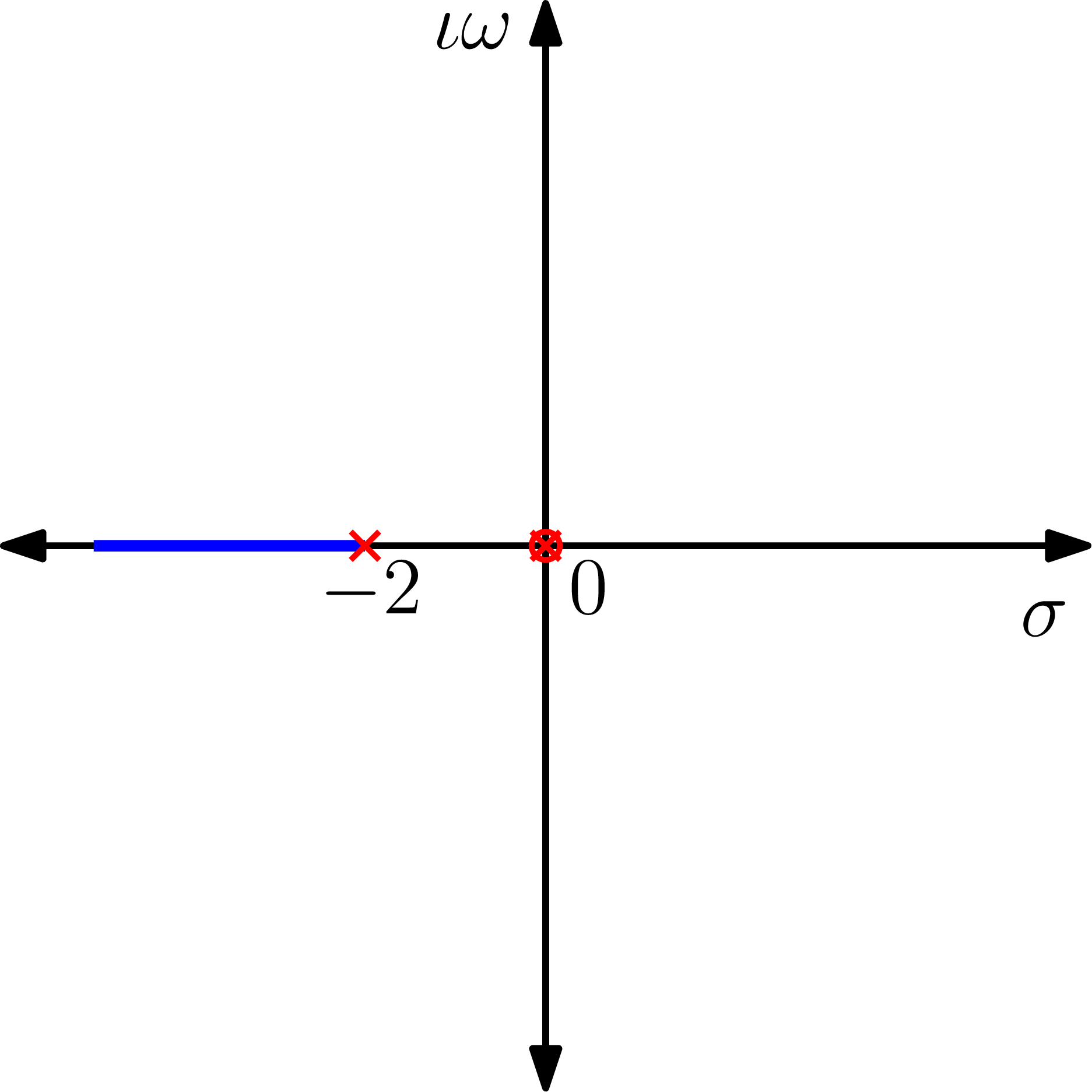}%
        \label{deflections}
    }
   \caption{Possible root loci with $k>0$ for different $\ell$.}
\label{complementary RL}
\end{figure}
From Fig. \ref{complementary RL}, it is evident that, for $k>0$, there is no imaginary axis crossing and all the non-zero eigenvalues of $-\mathcal{L}$ are in the left half plane for all $\ell =\{ 1,\dots,m\}$.

\textit{Case 2:} Here we consider the case when $k<0$. For simplicity, let us define $\bar{k}\coloneqq -k$. With this setup, equation \eqref{k_closed} can be rewritten as
\begin{equation} \label{k_rl}
    1-\bar{k} \underbrace{ \frac{s}{s^{2}+2s+ 2 \sin^{2}\left( \frac{\theta_{\ell}}{2} \right) +  2 \iota \sin\left( \frac{\theta_{\ell}}{2} \right) \cos \left( \frac{\theta_{\ell}}{2} \right)}}_{G_{\ell}(s)} =0.
\end{equation}
We will investigate how the zeros of $1- \bar{k}G_{\ell}(s)$ vary with changes in $\bar{k}$ for all $\ell=\{ 1,\dots,m\}$. Since \eqref{k_rl} is represented in general form and involves a polynomial of complex coefficients, we now consider the complementary root locus for a polynomial with complex coefficients as per \cite{doria2013root}.
As established earlier, $G_{\ell}(s)$ has its poles located in the left half of the complex plane and a zero at the origin. A sketch of the complementary root loci for different values of $\ell$ is shown in Fig. \ref{k_root locus}. We are interested in the range of gain $\bar{k}$ such that the root locus branches out only in the left half of the complex plane, ensuring all the non-zero eigenvalues of $-\mathcal{L}$ have negative real parts. Figure \ref{k_root locus} depicts that there is a branch of root locus going to the right half of the complex plane. 
\begin{figure}[t]
    \centering
    % Row 1
    \subfloat{%
        \includegraphics[width=0.4\linewidth]{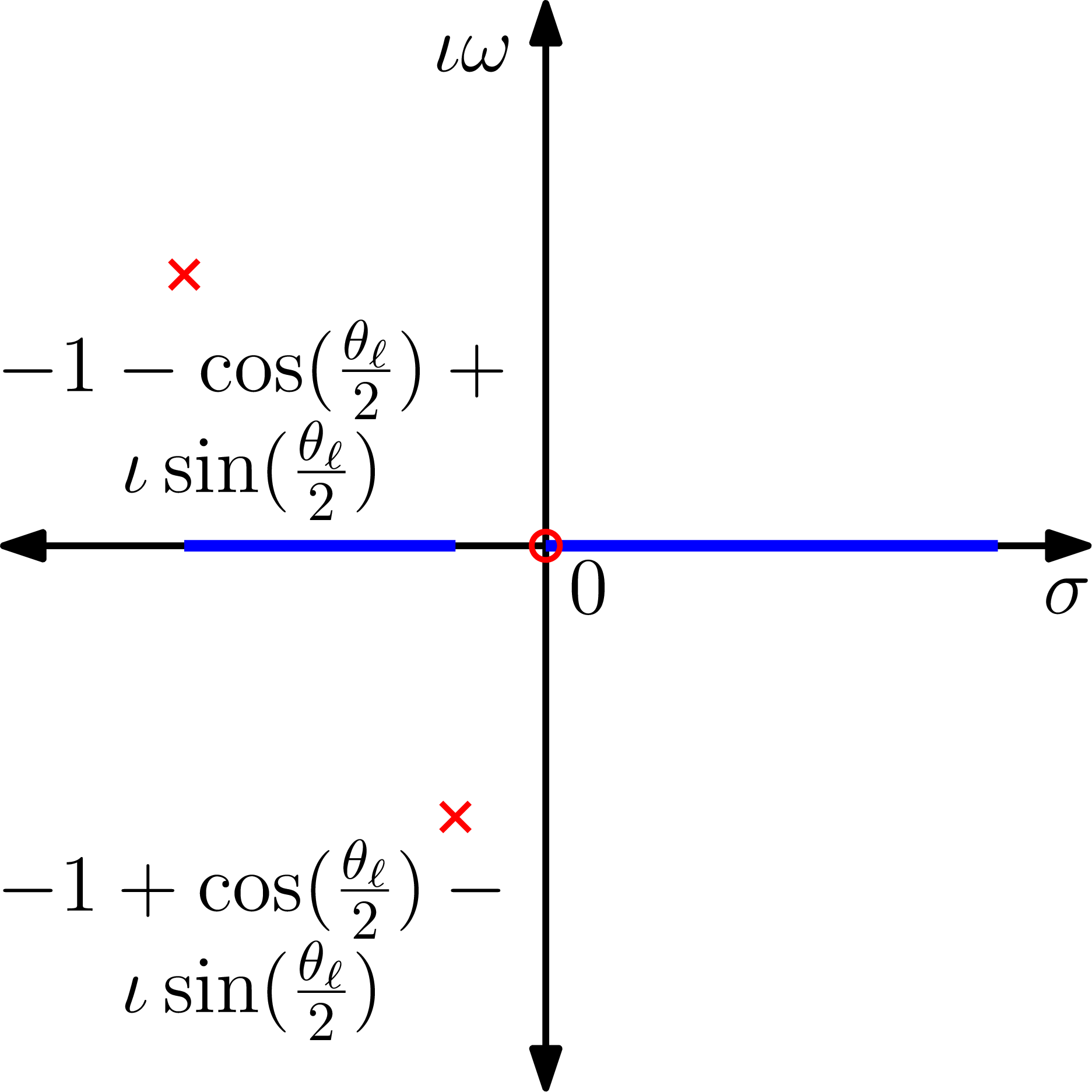}% 
    }\hfill
    \subfloat{%
        \includegraphics[width=0.4\linewidth]{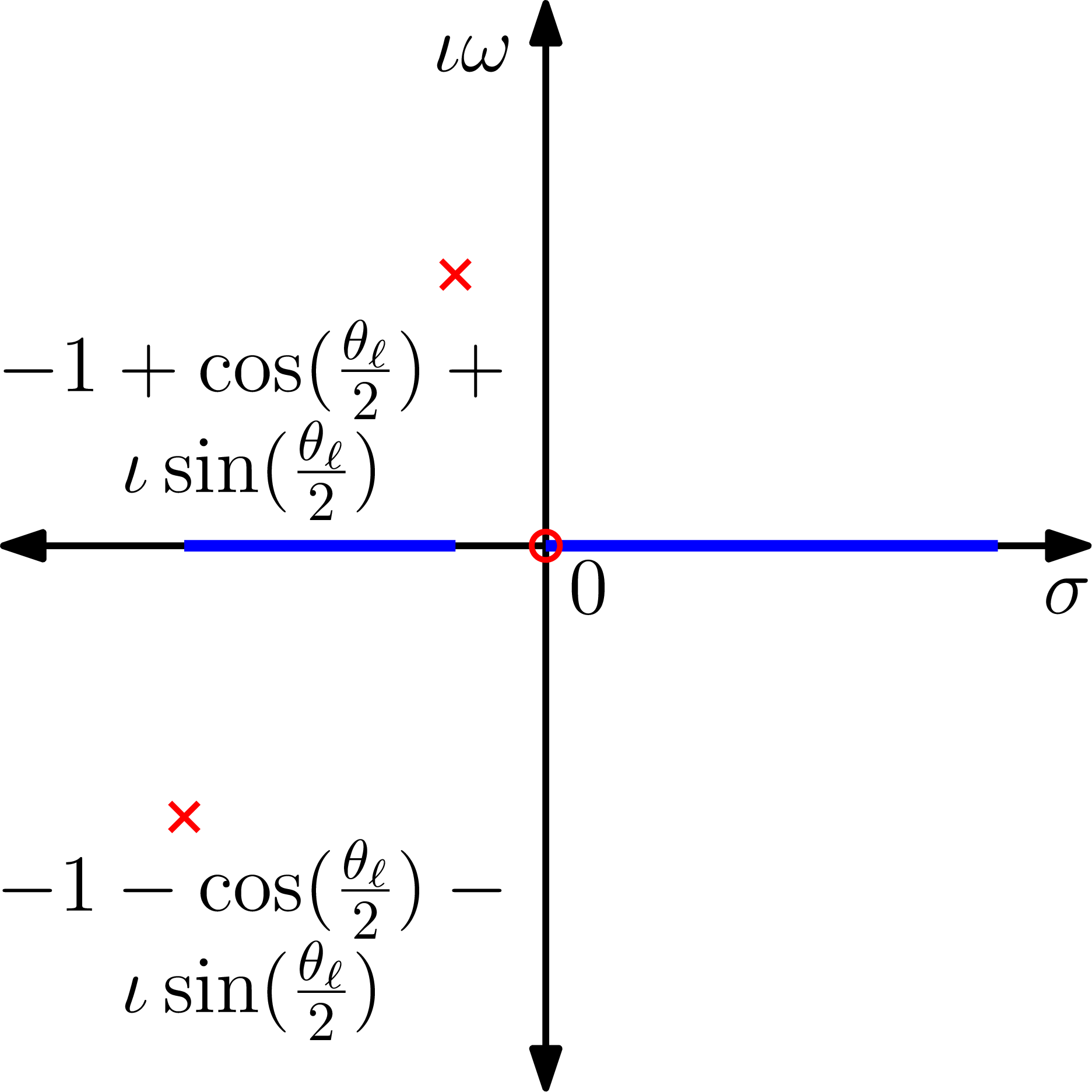}% 
    }\\[1ex]
    % Row 2
    \subfloat{%
        \includegraphics[width=0.4\linewidth]{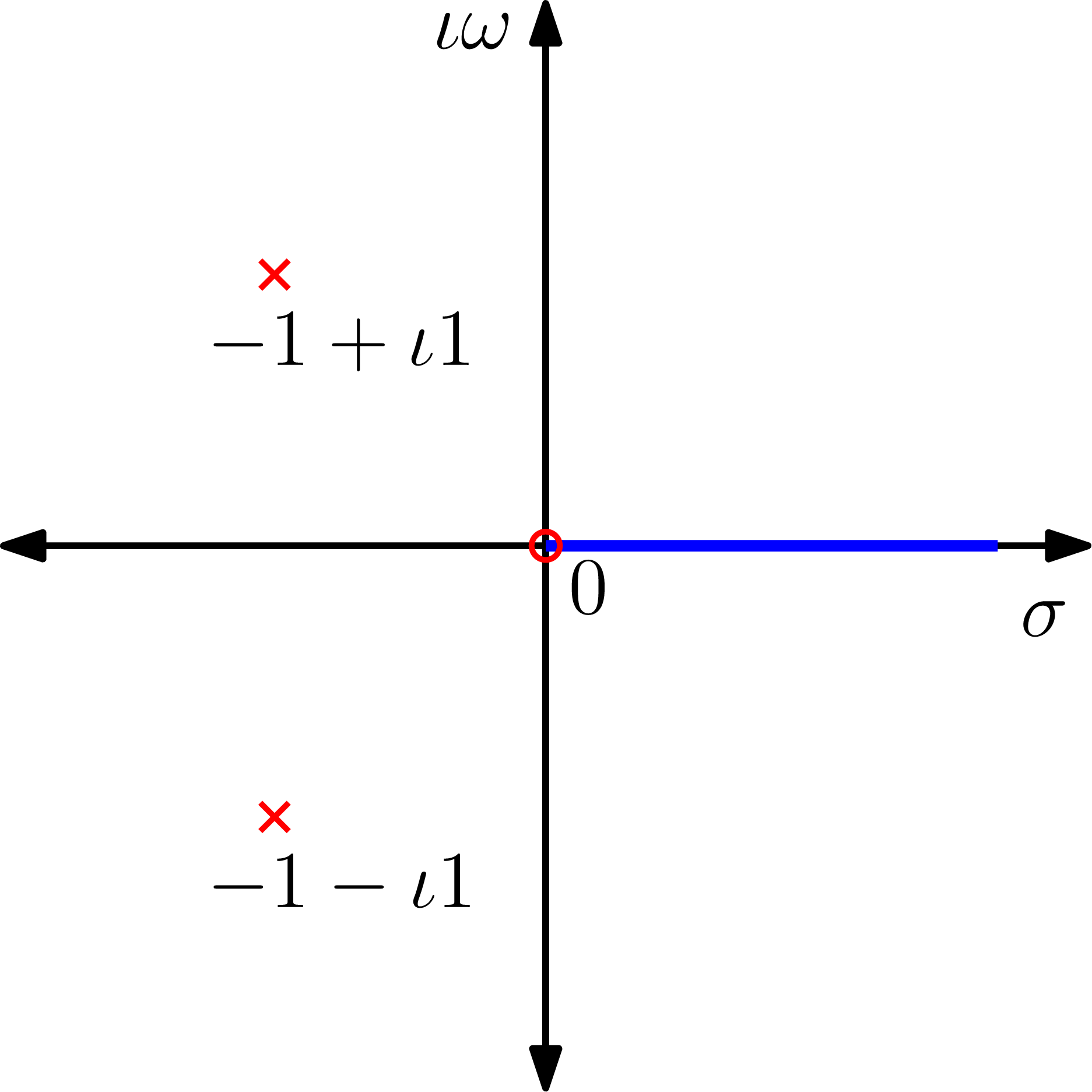}% 
    }\hfill
    \subfloat{%
        \includegraphics[width=0.4\linewidth]{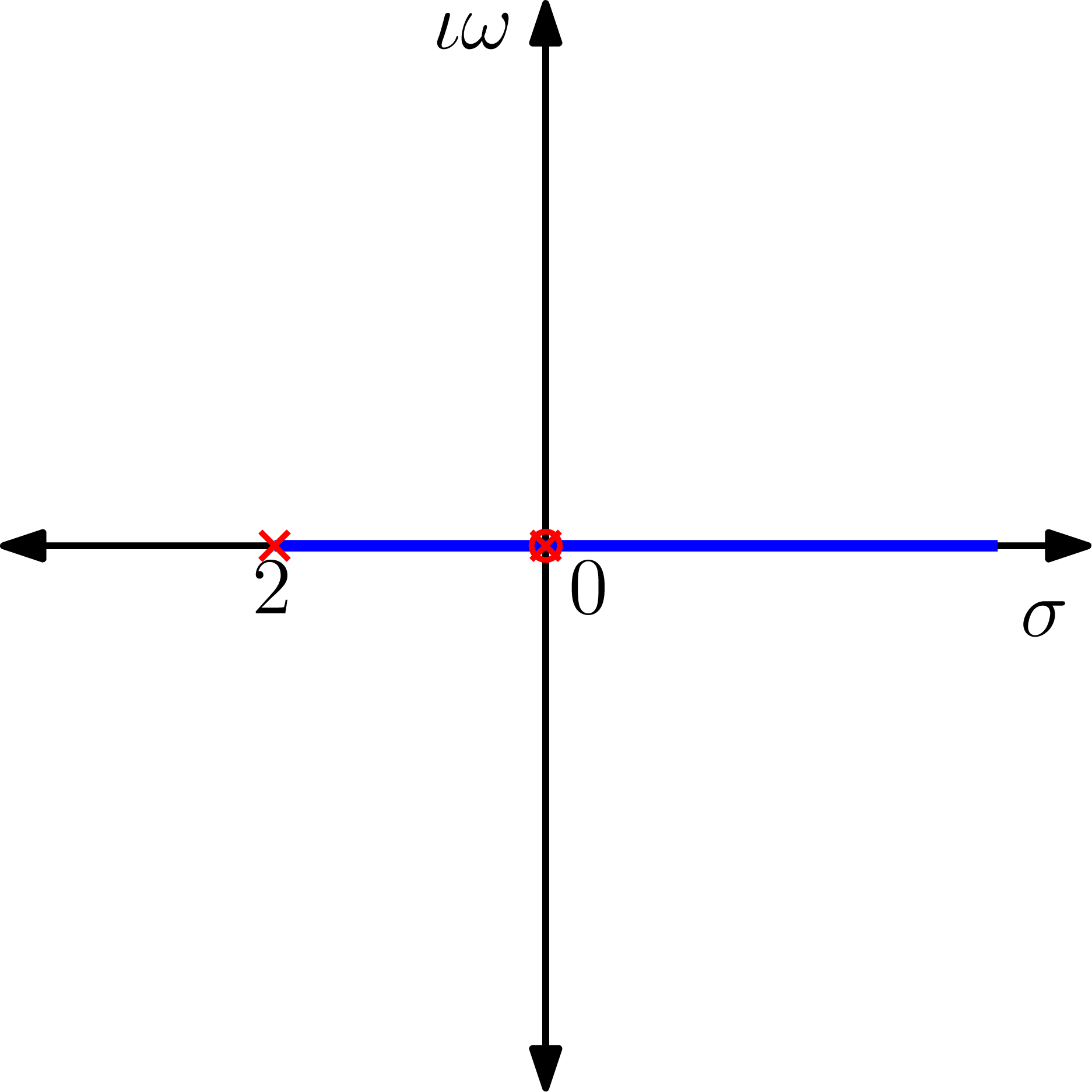}% 
    }
  \caption{Possible complementary root loci with $\bar{k}>0 ~(k<0)$ for different $\ell$.}
\label{k_root locus}
\end{figure}

For $\ell=1$, we have $\theta_1 =0$ and $G_1(s)$ has real poles. The closed loop polynomial would then be $s^2 + (2-\bar{k})s =0$. It is observed that at $\bar{k}=2$, the root locus is on the imaginary axis. Therefore, we must have $\bar{k} <2$ to ensure that the roots are located in the left half of the complex plane. Since we have to ensure all the non-zero eigenvalues of $-\mathcal{L}$ have negative real parts, we extend our analysis to other blocks, $\mathcal{D}_{\ell},~ \ell \neq 1$. 
Since we are interested in values of $\bar{k}$ for which the closed loop poles remain in the left half of the complex plane, we are essentially looking for the smallest gain (among all diagonal blocks of $\mathcal{D}$) at which any of the corresponding root loci branches crosses the imaginary axis for $\ell \in \{1,\dots,m\}$. 

To find out the imaginary crossing for $\ell=\{2,\dots,m\}$, we employ the complex Hurwitz test applicable to a polynomial with complex coefficients \cite{doria2013root}. Let a polynomial with complex coefficients be represented as $P(s)= s^2+ \delta_1 s + \delta_2 $; $\delta_i= a_i + \iota b_i$. The necessary and sufficient condition to have all its zeros in the left half plane is given by
\begin{align*} 
    \Delta_1 = a_1 >0,~~
    \Delta_2 = a_1^2a_2-b_2^2>0. 
\end{align*}
Applying these conditions to the polynomial $s^{2}+(2-\bar{k})s+ 2 \sin^{2}\left( \frac{\theta_{\ell}}{2} \right) + \iota 2\sin\left( \frac{\theta_{\ell}}{2} \right) \cos \left( \frac{\theta_{\ell}}{2} \right) =0$, we get the conditions
\begin{align*}
\begin{cases}
    2-\bar{k} >0 \\
   2 \left( 2-\bar{k}\right)^2 \sin^2 \left( \frac{\theta_{\ell}}{2} \right) - 4 \sin^2 \left( \frac{\theta_{\ell}}{2} \right) \cos^2 \left( \frac{\theta_{\ell}}{2} \right) >0, \\ \hspace{5cm}  \forall \ell =\{2,\dots,m\}
\end{cases} 
\end{align*}
Upon further simplification, one can get
\begin{align*}
    (2-\bar{k}) > \pm \frac{\sin \left(\theta_{\ell} \right)}{\sqrt{2} \sin \left(\frac{\theta_{\ell}}{2}\right)} = \pm \sqrt{2} \cos \left(\frac{\theta_{\ell}}{2}\right),
\end{align*}
\begin{align*}
    \implies \bar{k} < 2 \mp \sqrt{2} \cos \left(\frac{\pi (\ell -1)}{m}\right), \newline \forall \ell = \{2,\dots,m\}.
\end{align*}
Since we are looking for the smallest possible gain at the imaginary cross-over, this would result in 
\begin{align*}
     \bar{k} < 2 - \sqrt{2} \cos \left(\frac{\pi (\ell -1)}{m}\right), ~ \forall \ell = \{2,\dots,m\},
\end{align*}
\begin{align*}
    \implies \bar{k} < \min_{\ell} \left\{2 - \sqrt{2} \cos \left(\frac{\pi (\ell -1)}{m}\right)\right\}= 2 - \sqrt{2} \cos \left(\frac{\pi }{m}\right).
\end{align*}
This last inequality follows from the fact that cosine is a decreasing function in $\left(0, {\pi}/{2} \right)$. The bound $\bar{k} < 2 - \sqrt{2} \cos \left({\pi }/{m}\right)$ ensures all the roots of $G_{\ell}(s)$ are located in the left half of the complex plane. This indicates that $k$ should satisfy $k> -2 + \sqrt{2}\cos \left({\pi }/{m}\right) $ to have all the non-zero eigenvalues of $-\mathcal{L}$ in the left half of the complex plane, which in turn ensures $\mathcal{L}$ has all non-zero eigenvalues with positive real parts, thus completing the proof.
\end{proof}
\begin{rmk}\label{rem_block2}
    The bound on $k$, as mentioned in Theorem \ref{k_theorem}, indicates that the roots corresponding to block $\mathcal{D}_2$ determine the bound. This would also follow from observing that one of the poles of $G_{\ell}(s)$ is located at $-1+\cos \left(\frac{\theta_{\ell}}{2}\right)- \iota \sin \left(\frac{\theta_{\ell}}{2}\right)$, for all $\ell =\{ 2,\dots,m\}$. Since cosine is a decreasing function, the real part, $-1+ \cos \left(\frac{\theta_{\ell}}{2}\right)$, would decrease. The real part would have the least magnitude and thus be closest to the imaginary axis for $\ell=2$, irrespective of the number of macro-vertices. 
\end{rmk}
\begin{rmk}
    Theorem \ref{k_theorem} shows that the gain $k$ can admit negative values, thereby admitting antagonistic interactions. Unlike cycle pursuit, which can only have one edge with a negative weight \cite{sinha2006generalization}, our topology can permit $N/2$ negative edge weights out of $3N/2$ edges.
\end{rmk}

As we have ensured that all the non-zero eigenvalues of $\mathcal{L}$ are located in the right half of the complex plane, we have to decide on the values of $\alpha$ and $\beta$ to ensure closed-loop stability of the system in \eqref{overall eqn}, followed by Theorem \ref{yu_bound}.

\begin{figure}
    \centering
    \includegraphics[width=0.5\linewidth]{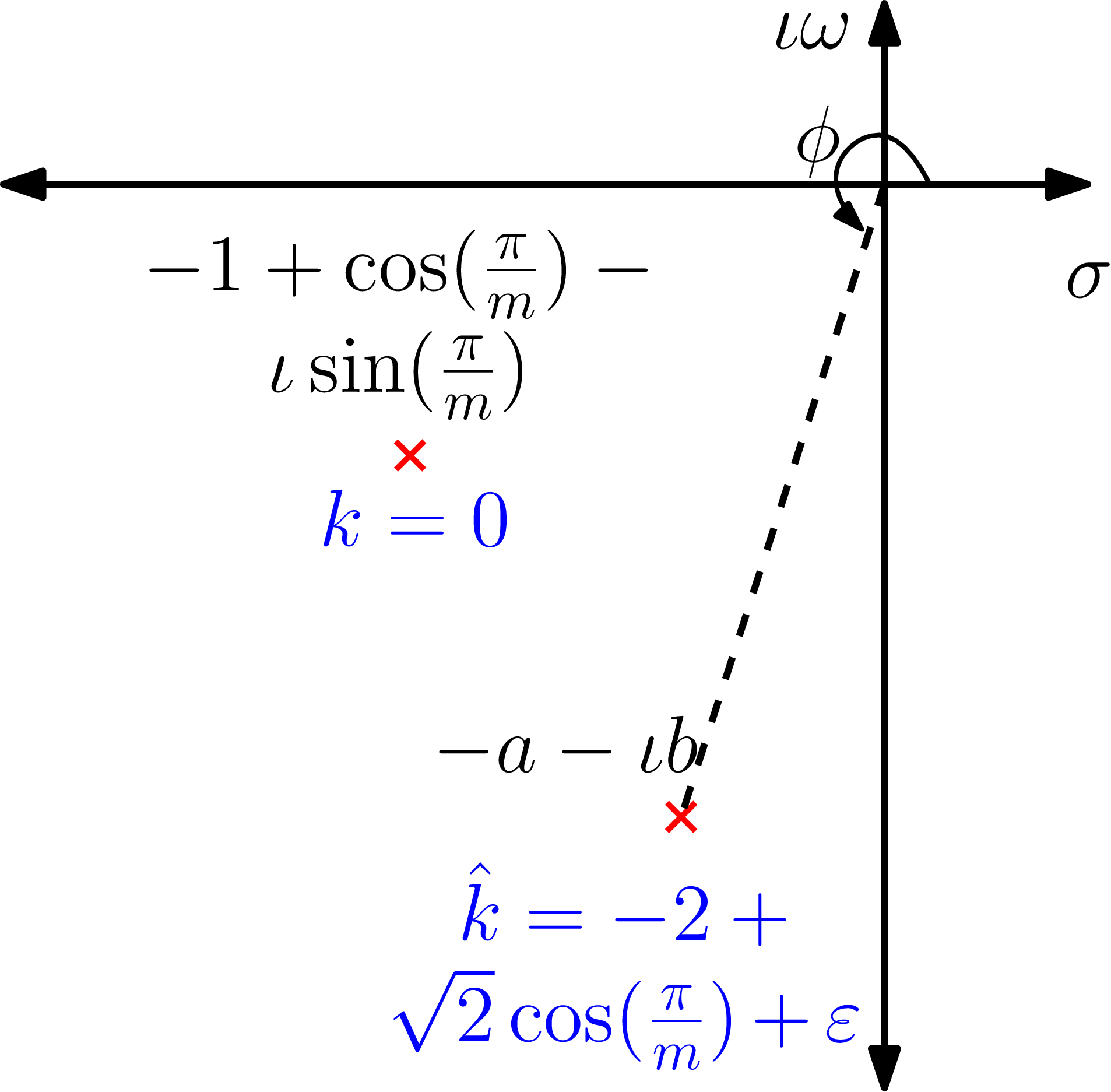}
    \caption{Illustration on design of $\alpha,\beta$ for a given $\hat{k}$.}
    \label{illustration}
\end{figure}

\begin{figure}
\centering
\begin{subfigure}{.25\textwidth}
  \centering
  \includegraphics[width=1.05\linewidth]{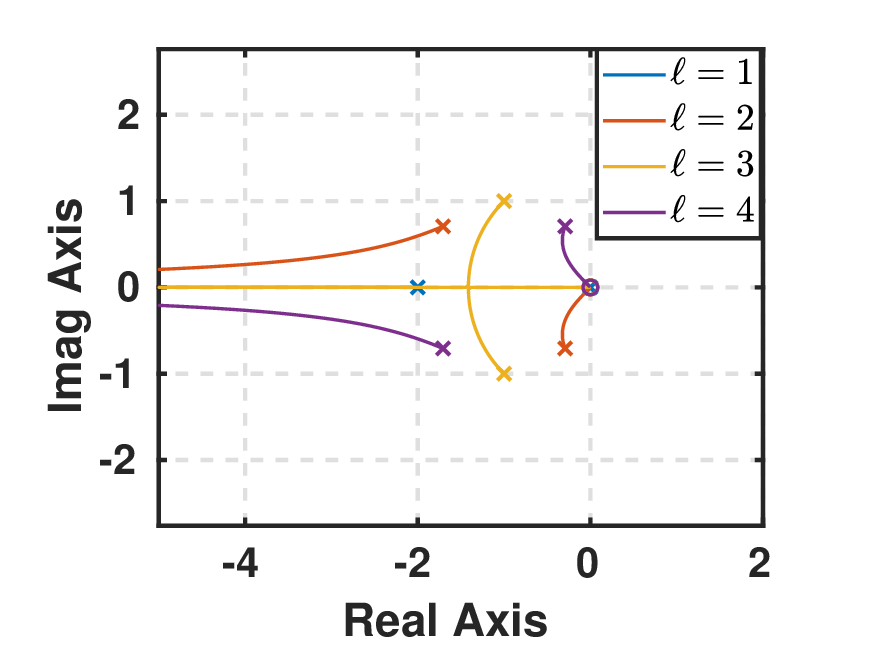}
  \caption{Root locus for $k>0$.}
  \label{root locus m4}
\end{subfigure}%
\begin{subfigure}{.25\textwidth}
  \centering
  \includegraphics[width=1.05\linewidth]{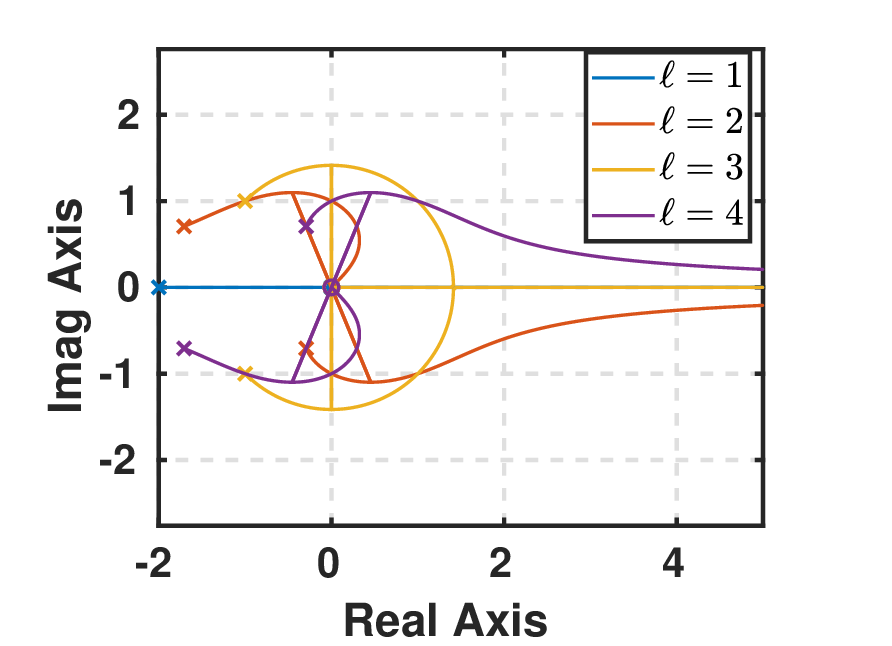}
  \caption{Complementary root locus, $k<0$. }
  \label{p_f sucess}
\end{subfigure}
\caption{Variation of $k$ (with $-\mathcal{L}$) for $m=4$.}
\label{complementary RL m4}
\end{figure}

We redefine the conditions on coupling factors from Theorem \ref{yu_bound} as
\begin{equation} \label{bound_a_b}
  \frac{\beta^{2}}{\alpha} >   \max_{2\le i\le N}\frac{\sin ^2 ({\phi}_i)}{|\mathfrak{Re}(\rho_i)|},
\end{equation}
where $\phi_i$ is the angle subtended in the anti-clockwise direction from the real axis to an eigenvalue $\rho_i$, with $\rho_i$ an eigenvalue of the $-\mathcal{L}$ and $|\mathfrak{Re}(\rho_i)|$ denotes the absolute value of $\mathfrak{Re}(\rho_i)$.
 This bound would be the maximum when $\sin(\phi_i)$ closer to $-1$ and  $\mathfrak{Re}(\rho_i)$ closer to zero.
From Remark \ref{rem_block2}, one of the eigenvalues of the block $\mathcal{D}_2$ has its real part closest to the imaginary axis, among all the non-zero eigenvalues of $-\mathcal{L}$. Let $\hat{k}$ be a gain closer to the imaginary axis (satisfying conditions as per Theorem \ref{k_root locus}), we evaluate the roots of $ P(s)=s^{2}+(2+\hat{k})s+ 2 \sin^{2}\left( \frac{\pi}{m} \right) + \iota 2\sin\left( \frac{\pi}{m} \right) \cos \left( \frac{\pi}{m}\right) =0$ (substituting $\ell=2$ for $\theta_{\ell}$) at this gain, $\hat{k}$. 
The roots of $P(s)$ are given by
\begin{align*}
    s= \frac{-2 -\hat{k} \pm \sqrt{\hat{k}^2 + 2\hat{k}+ 4\left(\cos \left( \frac{\pi}{m}\right)-\iota \sin \left( \frac{\pi}{m}\right)\right)}}{2}.
\end{align*}
Let the root of $P(s)$, which is closer to the imaginary axis, be given by $-a- \iota b$ for the given choice of $\hat{k}$. Therefore, the pole $-a-\iota b$ would make an angle $\phi$ with respect to the real axis as shown in Fig. \ref{illustration}. With this construction, we derive the conditions for $\alpha$ and $\beta$ such that $\frac{\beta^2}{\alpha} > \frac{\sin^2 \phi}{a}$.  
By this choice of construction, for a given control gain $k$, the designed $\alpha$ and $\beta$ corresponding to $\hat{k}$ would ensure robust stability for $k> \hat{k}$, indicating that the consensus among double integrators is achieved as per Theorem \ref{yu_bound} for the corresponding choices of $k$, $\alpha$ and $\beta$.
To better demonstrate this idea, let us consider the following example.
\begin{example}
    Consider a ring digraph with $4$ macro-vertices as shown in Fig. \ref{example m=4}. The relevant root locus and complementary root locus for variation in $k$ are shown in Fig. \ref{complementary RL m4}. The eigenvalues of block $\mathcal{D}_2$ are found to be $-0.2929-\iota 0.7071$ and $-1.7071+\iota 0.7071$. The bound on $k$ to ensure all the non-zero eigenvalues of $\mathcal{L}$ remain in the right half plane, as per Theorem \ref{k_theorem} is found to be $k> -1$. Let ${k}=-0.5$. We evaluate the polynomial $P(s)$ at this gain, which is given by $P(s)= s^2 +1.5 s+1 +\iota 1$. The roots of $P(s)$ are found to be $-0.1781-\iota 0.8744$ and $-1.3218 +\iota 0.8744$. Therefore, we need $\frac{\beta^2}{\alpha} > 5.39$. Let $\alpha=1$ and $\beta=6$. It can be verified that the matrix $\tilde{\mathcal{L}}$ in \eqref{closedloop} now has all its non-zero eigenvalues in the left half of the complex plane. It is observed that for this choice of $\alpha$ and $\beta$ values, any $k>-1$ ensures that all the non-zero eigenvalues of $\tilde{\mathcal{L}}$ are in the left half of the complex plane. For instance, if we choose $k=1$, the bound is calculated to be $\frac{\beta^2}{\alpha} > 2.13$. For $\alpha=1$ and $\beta=3$, stability of $\tilde{\mathcal{L}}$ is guaranteed. But, for the given choice of $\alpha=1$ and $\beta=3$, the stability of $\tilde{\mathcal{L}}$ is not ensured for $k=-0.8$, even if the nonzero eigenvalues of $\mathcal{L}$ are located in the left half of the complex plane. Choosing $k$ closer to the bound as per Theorem \ref{k_theorem}, and choosing $\alpha,\beta$ as per the bound as illustrated, would ensure consensus. 
\end{example}

\begin{rmk}
Note that deciding a single parameter, the control gain $k$ (satisfying Theorem \ref{k_theorem}), plays a vital role in ensuring stability of consensus and also in the choice of the design parameters $\alpha, \beta$ to guarantee the consensus of the network.  
\end{rmk}

Next, we analyze the common state at which consensus can be achieved.
\begin{theo}
Consider a group of double integrators driven by \eqref{closedloop} interacting over ring digraphs. Suppose the conditions on the control gain $k$ and design parameters $\alpha$ and $\beta$ are satisfied according to Theorem \ref{k_theorem} and \eqref{bound_a_b}. Let the initial positions and initial velocities be denoted as $p(0)\coloneqq\begin{bmatrix}
p_{1}(0) & p_{2}(0)  &\cdots   & p_{N}(0)
\end{bmatrix}^{T}$ and $v(0)\coloneqq\begin{bmatrix}
v_{1}(0) & v_{2}(0)  &\cdots   & v_{N}(0)
\end{bmatrix}^{T}$. The agents achieve consensus at a final velocity, $v_f$, given by
\begin{align}\label{final velocity}
    v_f = \frac{\sum_{i=1}^{m} (v_{2i-1}(0)+ (1+k)v_{2i}(0))}{m(2+k)}.
\end{align}
\end{theo}
\begin{proof}
The solution to \eqref{closedloop} is given by
\begin{align*}
    Z(t) = \sum_{i=3}^{2N} c_i \exp({\lambda_i t}) \xi_i + c_0 \begin{bmatrix}
        \mathbb{1}_N \\ 0_N
    \end{bmatrix} + c_1 \begin{bmatrix}
        \mathbb{1}_N t\\ \mathbb{1}_N
    \end{bmatrix},
\end{align*}
where $c_i \in \mathbb{C}$, $c_0,c_1 \in \mathbb{R}$, and $\xi_i \in \mathbb{C}^{2N}$ represent the eigenvectors corresponding to non-zero eigenvalues, $\lambda_i$, of $\tilde{\mathcal{L}}$.
For large values of $t$, we have $\exp^{\lambda_i t} \to 0$. This leads to $Z(t) \approx c_0 \begin{bmatrix}
        \mathbb{1}_N \\ 0_N
    \end{bmatrix} + c_1 \begin{bmatrix}
        \mathbb{1}_N t \\ \mathbb{1}_N
    \end{bmatrix}$.
Also, we have $\begin{bmatrix}
    0_N^T & w_2^T
\end{bmatrix} \dot{Z}(t) =0$, implying $w_2^T \dot{v}(t) =0$, where $w_2^T$ is the left eigenvector corresponding to the zero eigenvalue of $\mathcal{L}$, and is given by $w_2^T=\begin{bmatrix}
1 & 1+k & \cdots   & 1 & 1+k
\end{bmatrix}$. For all time $t$, we have, $\int_{0}^{t}w_2^{T}\dot{v}(\tau) d\tau=0$, which implies 
\begin{align*} w_2^{T}v(t)=w_2^{T}v(0)=\lim_{t\to\infty}w_2^{T}v(t) 
    \implies w_2^{T}v(0)= v_f w_2^{T}\mathbb{1}_N.
\end{align*}
Note that $v_f$ is the same as $c_1$.
This further simplifies to the following:
\begin{align*}
    v_f = \frac{w_2^T v(0)}{w_2^T \mathbb{1}_N} = \frac{w_2^T v(0)}{\sum_{i=1}^N w_{2_{i}}}
\end{align*}
which, combined with the observation that $\sum_{i=1}^N w_{2_{i}}=m(2+k)$, completes the proof.
\end{proof}

%%%%%%%%%%%%%%%%%%%%%%%%%%%%%%%%
\subsection{Application to a swarm of quadrotors}
In this section, we focus on adapting the results for double integrators to the dynamics in \eqref{position dynamics} to achieve the desired formation of a swarm of quadrotors interacting over ring digraphs, thereby enabling them to move with a common flight velocity. The objective is to achieve a desired formation at a specified altitude. Due to low rotational inertia and high torque, the time constant of the position dynamics is larger compared to that of the attitude dynamics. This approximation allows the three control inputs to be designed independently, to control in all different directions, and is widely accepted in the literature \cite{dong2014time}. 

Thus, the control law in \eqref{control} is modified as
\begin{equation}
\begin{split}
       u_{i\ell} (t)&= - \alpha\sum_{j \in \mathcal{N}_i}^{} a_{ij} \left( p_{\ell i}(t)-p_{\ell j}(t) -d_{ij}^* \right)  \\
       &- \beta \sum_{j \in \mathcal{N}_i}^{} a_{ij} \left( v_{\ell i}(t)-v_{\ell j}(t) \right), ~\ell = x,y\\
       u_{iz}(t) &= k_{pz} \left( p_{zi}(t)-z_{\text{com}} \right) + k_{vz} v_{zi}(t) 
       \end{split}
\end{equation}
where $d_{ij}^*$ represents the desired formation shape in the horizontal plane specified by the user, $k_{pz}$ and $k_{vz}$ denote the proportional and derivative gains, $z_{\text{com}}$ represents the desired altitude to be achieved by the swarm, $p_{\ell i}(t)$ and $v_{\ell i}(t)$, for $\ell= x,y,z$ denote the position and velocity of $i^{\text{th}}$ quadrotor in respective directions. Although the modified control law has a constant term, $d_{ij}^*$, it does not vary much from the consensus dynamics. 

The desired thrust, the reference roll, pitch, and yaw are as follows:
\begin{align*}
\begin{split}
    T_i &= m_i \sqrt{u_{ix}^2 +u_{iy}^2+ (u_{iz}+g)^2} \\
\phi_{id} & = \arcsin \left( \frac{m_i}{T_i} \left( u_{ix} \sin\psi_{id} - u_{iy} \cos\psi_{id}\right) \right) \\
\theta_{id} &= \arctan \left( \frac{1}{u_{iz}+g} \left( u_{ix} \cos \psi_{id} + u_{iy} \sin\psi_{id} \right) \right) \\
\psi_{id} &=0
\end{split}
\end{align*}
In the attitude loop, traditional proportional-integral-derivative controllers are used to track the desired references $\phi_{id}$, $\theta_{id}$, and $\psi_{id}$, using suitable gains \cite{wang2024controllers}. Note that one can also use any other controller to track the desired values of thrust and attitude of quadrotors.
%%%%%%%%%%%%%%%%%%%%%%%%%%%%%%%5
\subsection{Analysis of achievable velocities}
\begin{figure}[!h]
    \centering
\begin{subfigure}{0.25\textwidth}
      \centering
  \includegraphics[width=0.8\linewidth]{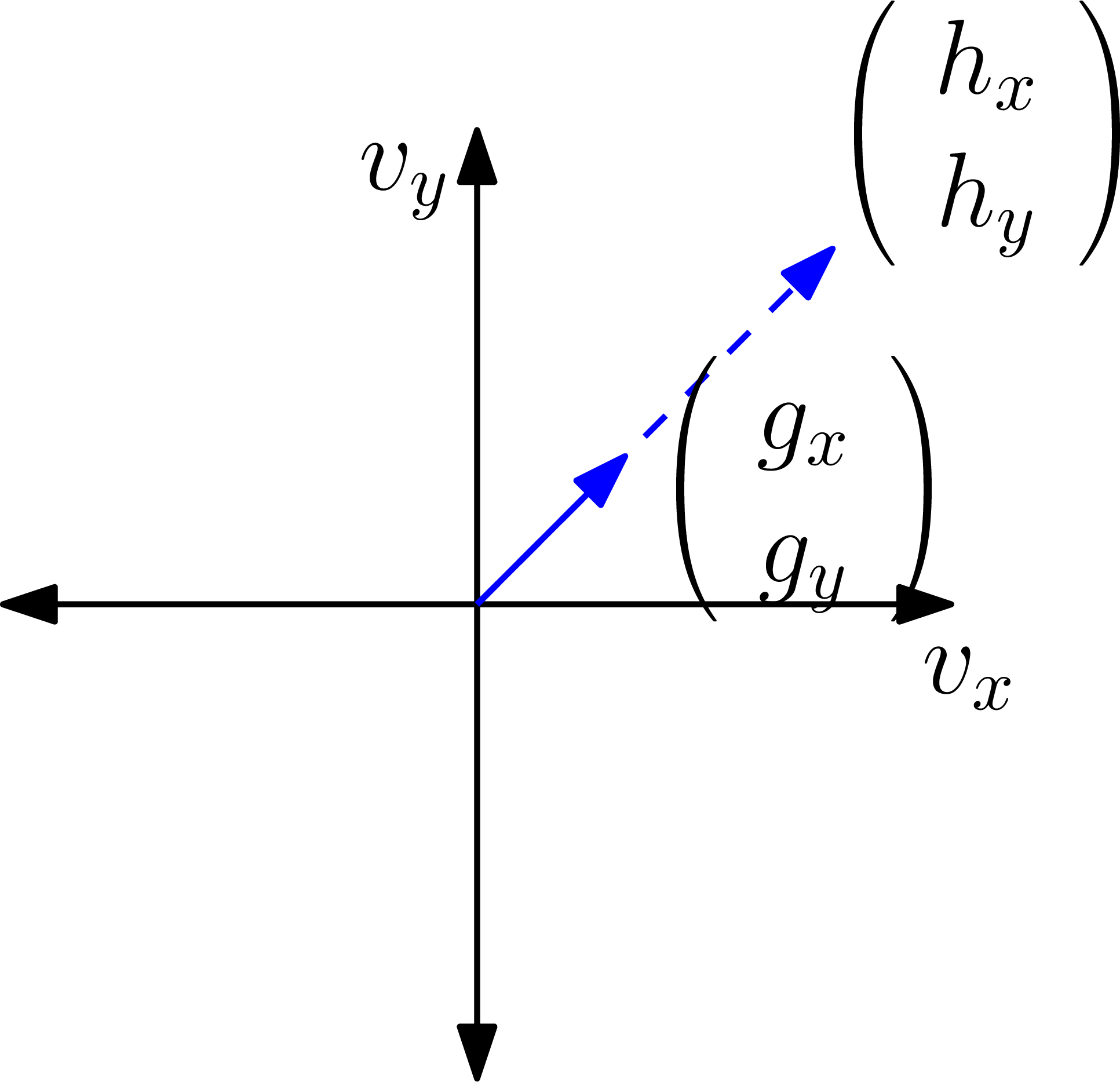}
  \caption{Positive scaling.}
 % \label{root locus m4}
\end{subfigure}%
\begin{subfigure}{.25\textwidth}
  \centering
  \includegraphics[width=0.8\linewidth]{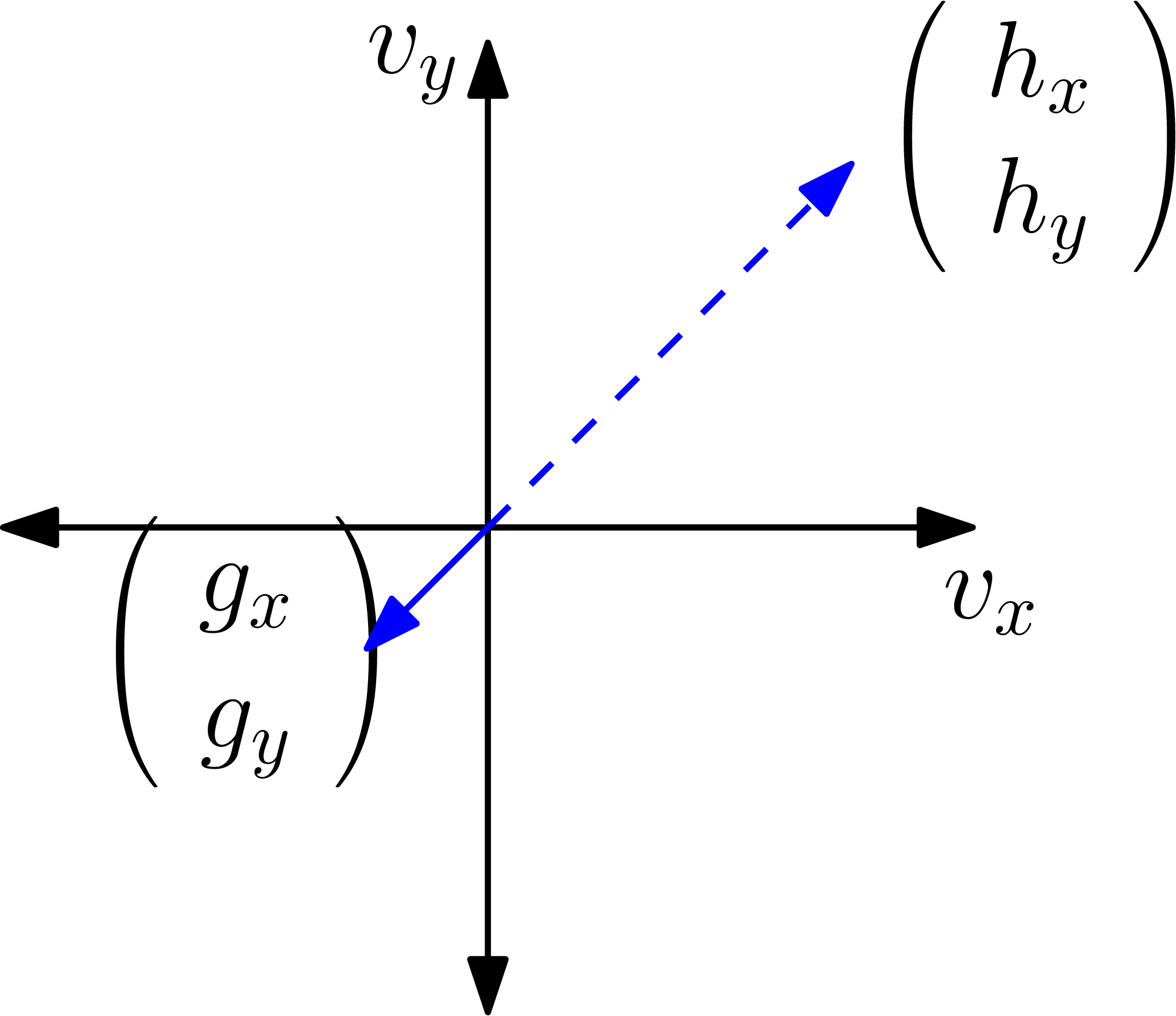}
  \caption{Negative scaling.}
  %\label{p_f sucess}
\end{subfigure}
    \caption{Illustration of \eqref{linear_comb}.}
    \label{achivable velocity}
\end{figure}

Suppose we consider the position of the agents in $\mathbb{R}^2$. With this setup, a given final velocity in $\mathbb{R}^2$ is represented as $v_f = v_{xf} +\iota v_{yf}$ and given initial velocities are defined as $v_i(0) = v_{xi} (0) +\iota v_{yi} (0)$.
We are interested in deriving a set of achievable velocities through the choice of control gain, $k$, starting from arbitrary initial velocities. Expanding \eqref{final velocity}, we get
\begin{align*}
    \sum_{i=1}^{m} (2+k) v_f = \sum_{i=1}^{m} v_{2i-1}(0)+ \sum_{i=1}^{m} (1+k)v_{2i}(0)
\end{align*}
\begin{align*}
    \implies (1+k)  \sum_{i=1}^{m} (v_f-v_{2i}(0))= \sum_{i=1}^{m} (v_{2i-1}(0)-v_f)
\end{align*}
Suppose we define $g_x \coloneqq \sum_{i=1}^{m}(v_{xf}-v_{x_{2i}}(0))$,~  $h_x\coloneqq \sum_{i=1}^{m} (v_{x_{2i-1}}(0)-v_{xf})$, $g_y\coloneqq\sum_{i=1}^{m}(v_{yf}-v_{y_{2i}}(0))$ and $h_y\coloneqq \sum_{i=1}^{m} (v_{y_{2i-1}}(0)-v_{yf})$. Then, these equations can be represented as a point in the $(v_x,v_y)$ plane as
\begin{align} \label{linear_comb}
  \delta  \begin{bmatrix}
g_x \\
g_y
\end{bmatrix}= \begin{bmatrix}
h_x \\
h_y
\end{bmatrix},
\end{align}
where $\delta=(1+k)$. Geometrically, equation \eqref{linear_comb} can be viewed as one vector being a scalar multiple of the other, as in Fig. \ref{achivable velocity}.
Any value of scaling $\delta$ may not be feasible as $k$ has to satisfy the bound as per Theorem \ref{k_theorem}. Given a set of initial velocities, the desired final velocity in the $(v_x,v_y)$ plane is achievable if it is possible to express $\begin{bmatrix}
g_x &
g_y
\end{bmatrix}^T$ as a scalar multiple of $\begin{bmatrix}
h_x &
h_y
\end{bmatrix}^T$ such that $k$ satisfies the bound as per Theorem \ref{k_theorem}.

\begin{figure}
    \centering
    \includegraphics[width=0.5\linewidth]{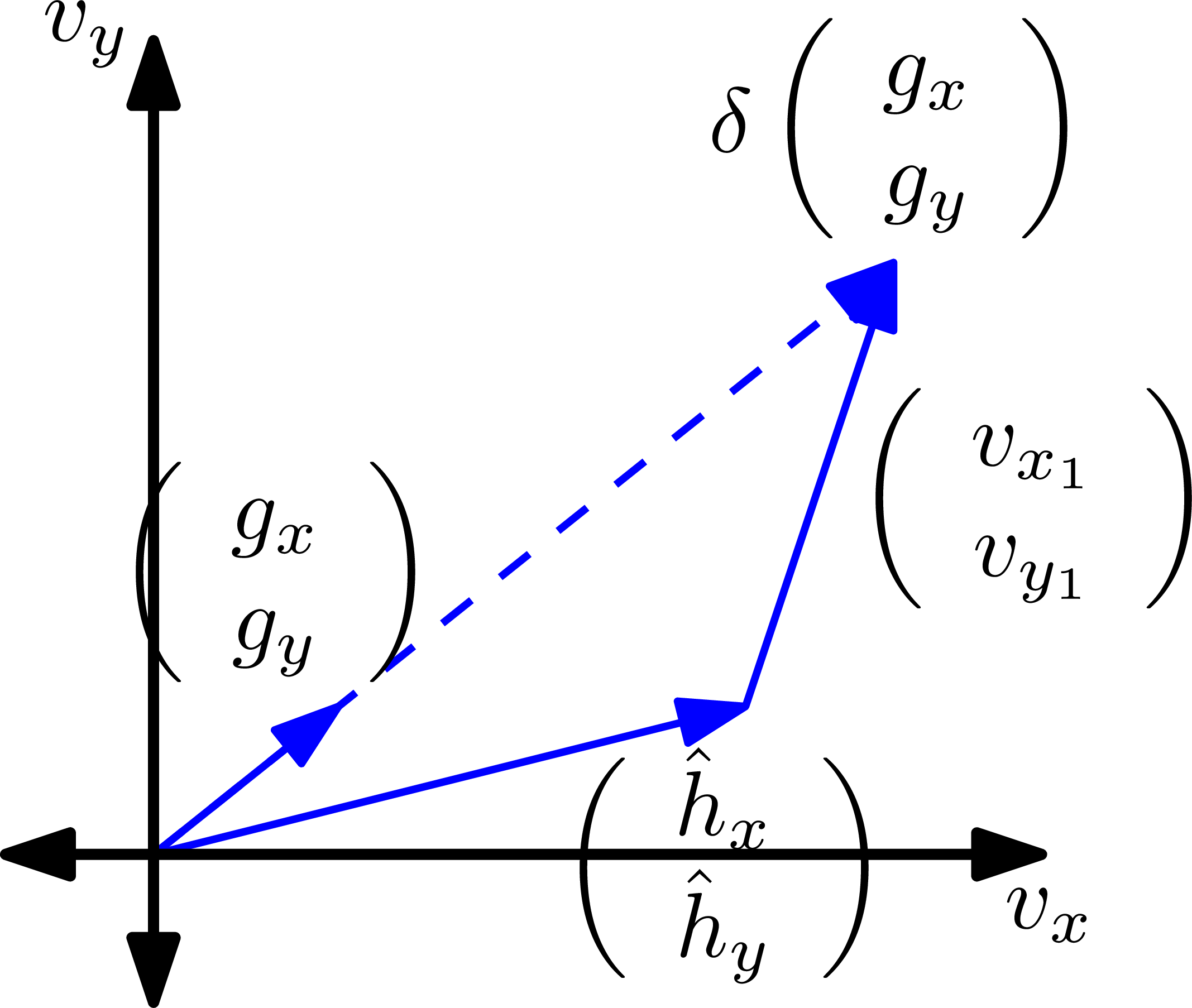}
    \caption{Illustration of \eqref{linear_comb_modified}.}
    \label{achie_modified}
\end{figure}

Suppose the initial velocities and the desired final velocity are such that their linear combinations are not representable as in \eqref{linear_comb}. Then it is possible to achieve the desired final velocity by modifying the initial velocity of any one of the odd-indexed agents such that \eqref{linear_comb} is satisfied. Without loss of generality, let the initial velocity of the first agent, i.e., $i=1$, be chosen. Suppose we define $\hat{h}_x\coloneqq \sum_{i=2}^{m} \left (v_{x_{2i-1}}(0)-v_{xf} \right) - v_{xf}$ and $\hat{h}_y\coloneqq \sum_{i=2}^{m} \left (v_{y_{2i-1}}(0)-v_{yf} \right) - v_{yf}$. Then, equation \eqref{linear_comb} may be rewritten as
\begin{align} \label{linear_comb_modified}
      \delta  \begin{bmatrix}
g_x \\
g_y
\end{bmatrix}= \begin{bmatrix}
\hat{h}_x \\
\hat{h}_y
\end{bmatrix} + \begin{bmatrix}
v_{x_1} \\
v_{y_1}
\end{bmatrix}.
\end{align}
Geometrically, the modified equation \eqref{linear_comb_modified} may be illustrated as in Fig. \ref{achie_modified}.

Observe that, depending on the choice of control gain $k$ (satisfying Theorem \ref{k_theorem}), we may choose $\delta$ and $\begin{bmatrix}
v_{x_1} &
v_{y_1}
\end{bmatrix}^T$ such that \eqref{linear_comb_modified} holds. 
The selection of the first agent for initial velocity assignment is not unique and could be applied to any of the odd-indexed agents. Nevertheless, the choice of $\delta$ would vary, indicating that there are numerous admissible choices of $k$ that yield the desired final velocity. This concept is illustrated through the following example.
\begin{example}
    Consider the ring digraph with $m=3$. Suppose the initial velocities of the agents are given by $v_1= \begin{bmatrix}
    1 & 2 \end{bmatrix} ^T$, $v_2= \begin{bmatrix}
    -1 & 3 \end{bmatrix} ^T$, $v_3= \begin{bmatrix}
    -2 & 4 \end{bmatrix} ^T$, $v_4= \begin{bmatrix}
    1 & 5 \end{bmatrix} ^T$, $v_5= \begin{bmatrix}
    2 & 3 \end{bmatrix} ^T$ and $v_6= \begin{bmatrix}
    2 & 2 \end{bmatrix} ^T$. Let the desired final velocity be $v_f= \begin{bmatrix}
    5 & -4 \end{bmatrix} ^T$. Clearly, $h=\begin{bmatrix}
    -14 & 20 \end{bmatrix} ^T$ cannot be expressed as scaled version of $g= \begin{bmatrix}
    13 & -23 \end{bmatrix} ^T$. If we take $\delta=1.5$, then \eqref{linear_comb_modified} may be written as
    \begin{align}\label{modified_v1}
        1.5 \begin{bmatrix}
13 \\
-23
\end{bmatrix}= \begin{bmatrix}
-15 \\
18
\end{bmatrix} +\underbrace{ \begin{bmatrix}
v_{x_1} \\
v_{y_1} \end{bmatrix}}_{v_1},
    \end{align}
and solving for $v_1$, we get a new initial velocity $v_1$ as $v_1= \begin{bmatrix}
    34.5 & -52.5 \end{bmatrix} ^T$. With this $v_1$, we rewrite \eqref{linear_comb} as
    \begin{align*}
        \delta \begin{bmatrix}
13 \\
-23
\end{bmatrix}= \begin{bmatrix}
-19.5 \\
-34.5
\end{bmatrix} .
    \end{align*}
The gain $\delta$ is evaluated to be $\delta=1.5$, thereby leading to $k=0.5$, which satisfies Theorem \ref{k_theorem}. With $k=0.5$, the desired final velocity is achieved if the first agent modifies its initial velocity accordingly.

Instead, if the third agent's initial velocity is modified, then equation \eqref{linear_comb_modified} would become
   \begin{align*}
        0.5 \begin{bmatrix}
13 \\
-23
\end{bmatrix}= \begin{bmatrix}
-12 \\
16
\end{bmatrix} + \underbrace{\begin{bmatrix}
v_{x_3} \\
v_{y_3} \end{bmatrix}}_{v_3},
    \end{align*}
where $\delta$ is chosen as $\delta=0.5$. The modified velocity for $v_3$ is found to be $v_3= \begin{bmatrix}
    18.5 & -27.5 \end{bmatrix} ^T$. With this new velocity, the desired final velocity is achieved with the control gain of $k=-0.5$, satisfying the conditions as per Theorem \ref{k_theorem}.
\end{example}
%%%%%%%%%%%%%%%%%%%%%%%%%%%%%%%%
\begin{rmk}
    By selecting a single control gain, $k$, we can characterize the entire set of final velocities that the system can attain for a given set of initial velocities. This clearly explains that changes in the single control gain $k$ can effectively influence the closed-loop stability, consensus value, and the final flight velocity.
\end{rmk}
\begin{rmk}
    Note that the idea of modifying the velocity of one agent, as in \eqref{modified_v1}, is similar to the one in the two-stage algorithm proposed in \cite{mukherjee2013reachability} with $N+6$ variables. However, our method achieves the two-stage modification using only a single control gain $k$. In addition, the control law in \cite{mukherjee2013reachability} ensured positional consensus with zero velocity. In contrast, our problem focuses on achieving a desired shape and thereafter moving along a desired direction with a non-zero velocity.
\end{rmk}
%%%%%%%%%%%%%%%%%%%%%%%%%%%%%%%%%%%%
\section{SIMULATION STUDIES}
For simulations, we consider $6$ quadrotors interacting over a ring digraph with $m=3$. The initial positions of each quadrotor are given by $p_1(0)=\begin{bmatrix} 0 & 20 &0 \end{bmatrix}^T$, $p_2(0)=\begin{bmatrix} 50 & -20 &0 \end{bmatrix}^T$, $p_3(0)=\begin{bmatrix} 10 & 30 &0 \end{bmatrix}^T$, $p_4(0)=\begin{bmatrix} 10 & 20 &0 \end{bmatrix}^T$, $p_5(0)=\begin{bmatrix} 20 & 0 &0 \end{bmatrix}^T$ and $p_6(0)=\begin{bmatrix} 20 & 10 &0 \end{bmatrix}^T$. The initial velocities were $v_1(0)=\begin{bmatrix} 9 &4& 0 \end{bmatrix}^T$, $v_2(0)=\begin{bmatrix} 15 &8& 7 \end{bmatrix}^T$, $v_3(0)=\begin{bmatrix} 18 &-2& 13 \end{bmatrix}^T$, $v_4(0)=\begin{bmatrix} 13 &1& 7 \end{bmatrix}^T$, $v_5(0)=\begin{bmatrix} -8 &-7& 6 \end{bmatrix}^T$, and $v_6(0)=\begin{bmatrix} 14 &18& -6 \end{bmatrix}^T$. The desired formation is chosen to be a hexagon inscribed inside a circle of radius $20$m with each vertex of the hexagon at angles $\theta = \frac{2 \pi (\ell -1)}{6}, \ell=1,\dots,6$. It is desired that a hexagon formation at a desired altitude of $z=50$ m is achieved with flight velocity $v_f$ in the horizontal plane. The control gain $k$ is given by $k=5$, and the values of $\alpha$ and $\beta$ are chosen as $\alpha=1$ and $\beta=5$, satisfying the stability conditions. Simulations are conducted for these parameters, and a set of stabilizing gains is determined to track the desired attitude angles using our proposed protocol.

\begin{figure}
    \centering
    \includegraphics[width=0.8\linewidth]{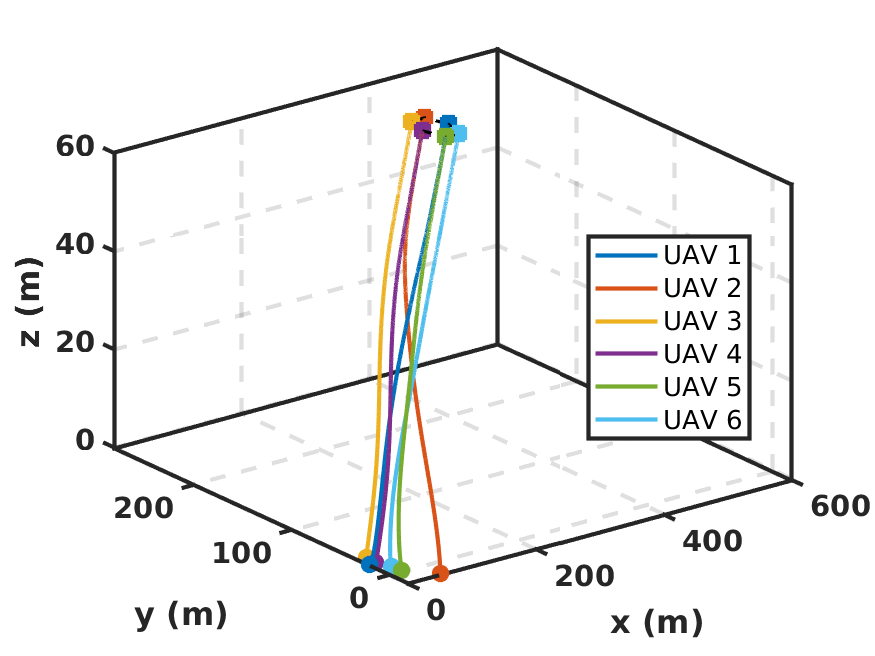}
    \caption{$3$D trajectories of the quadrotor UAVs.}
    \label{3dplot}
\end{figure}

\begin{figure}
    \centering
\begin{subfigure}{0.25\textwidth}
      \centering
  \includegraphics[width=1.1\linewidth]{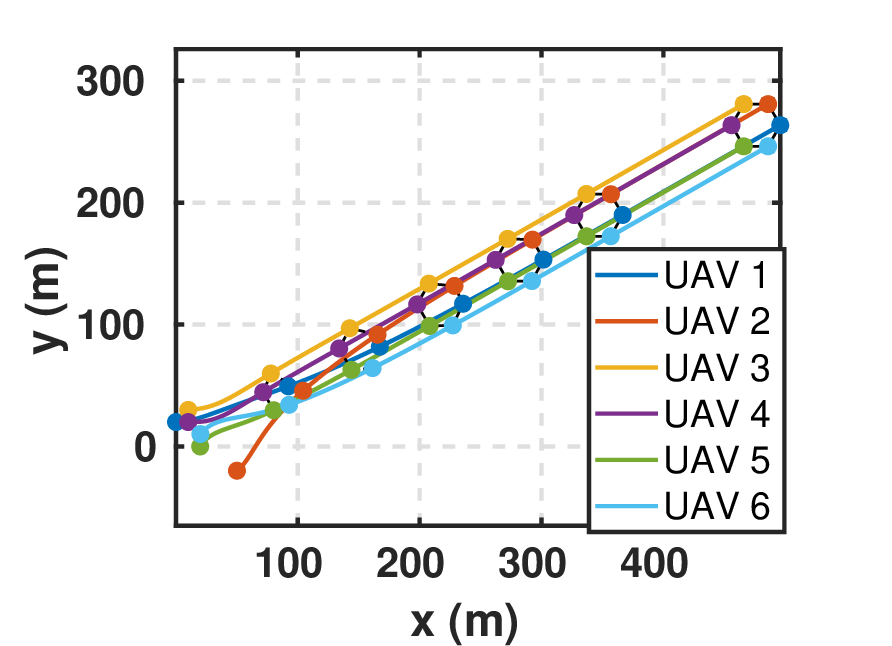}
  \caption{Trajectories of desired formation.}
 % \label{root locus m4}
\end{subfigure}%
\begin{subfigure}{.25\textwidth}
  \centering
  \includegraphics[width=1.1\linewidth]{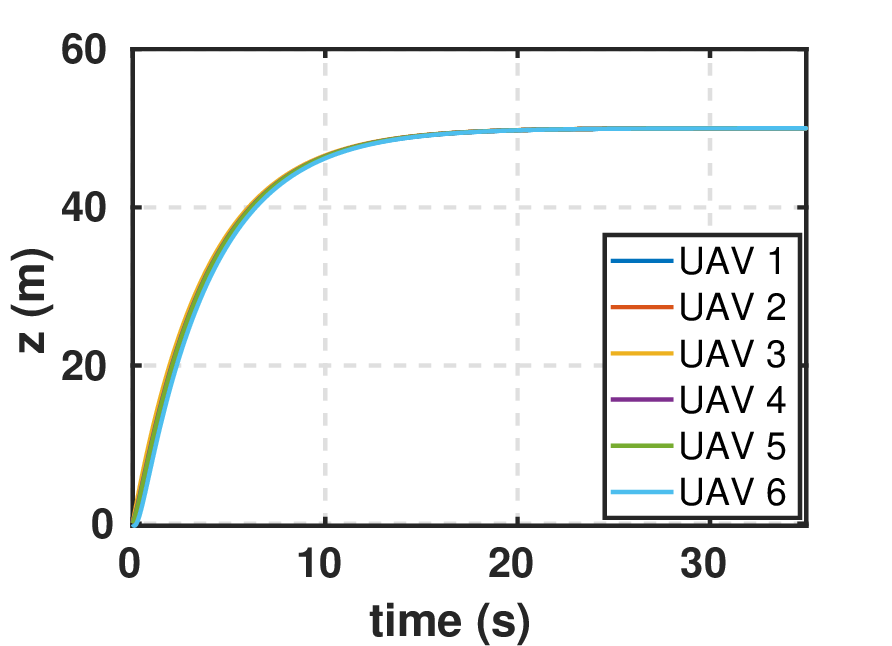}
  \caption{Altitude trajectories. }
  %\label{p_f sucess}
\end{subfigure}
    \caption{Desired formation at specified height.}
    \label{formation and height}
\end{figure}

\begin{figure}
    \centering
    \includegraphics[width=0.8\linewidth]{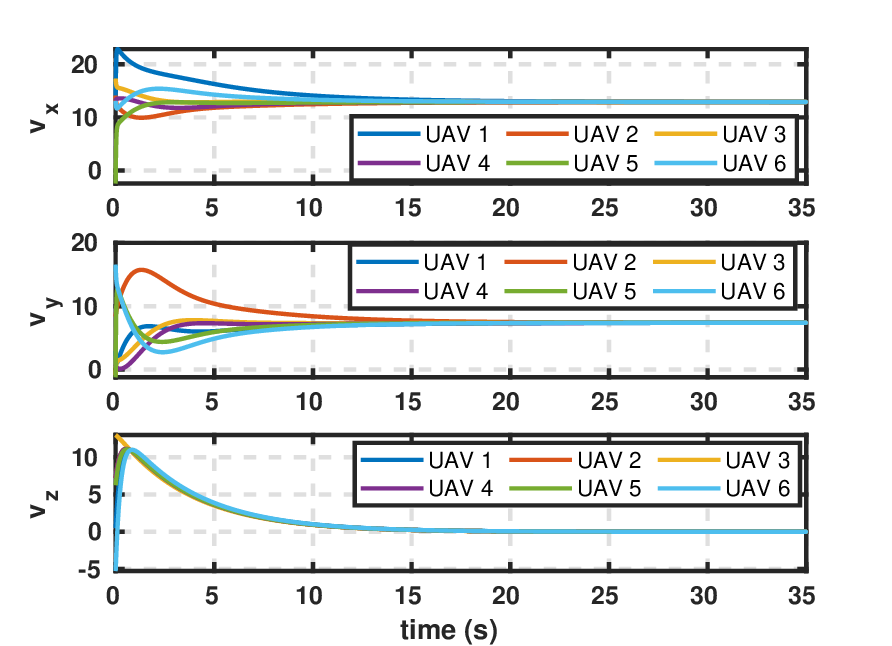}
    \caption{Velocity profiles of the quadrotor UAVs.}
    \label{veloity}
\end{figure}

\begin{figure}
    \centering
    \includegraphics[width=0.8\linewidth]{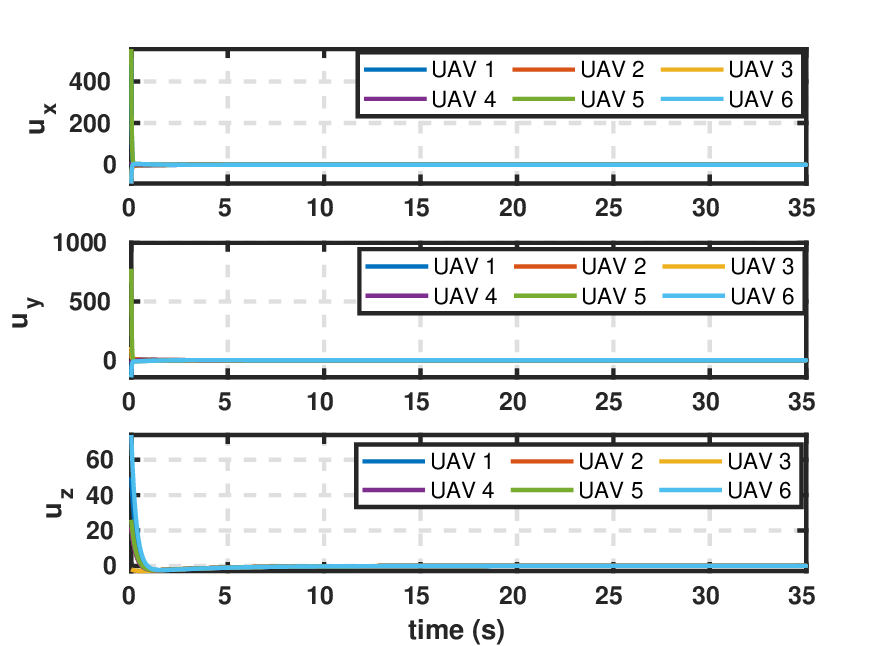}
    \caption{Amplitude profiles of the quadrotor UAVs.}
    \label{amplitude}
\end{figure}

\begin{figure}
    \centering
    \includegraphics[width=0.8\linewidth]{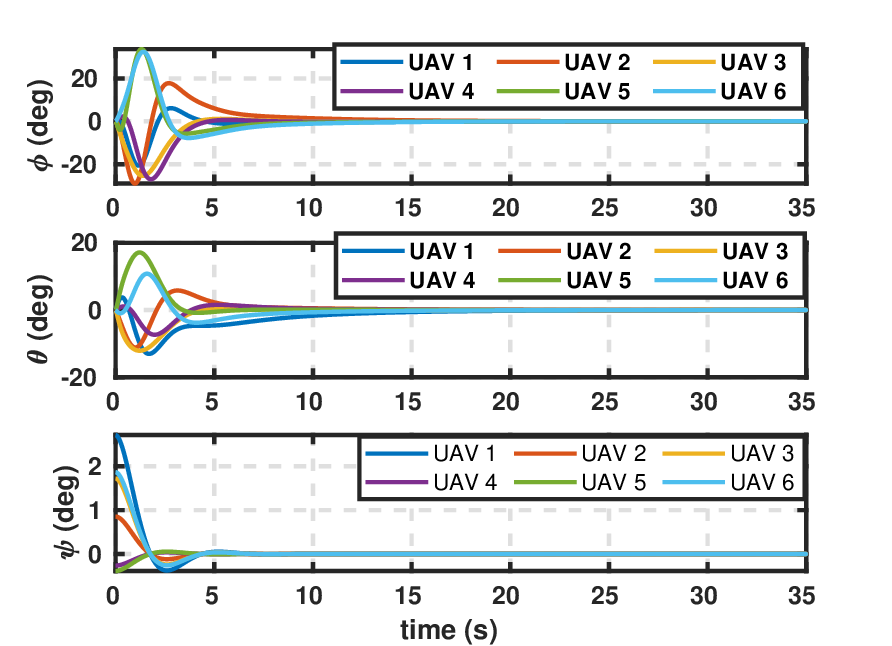}
    \caption{Attitude angles of UAVs.}
    \label{phi_theta_psi}
\end{figure}

The trajectories of UAVs in three-dimensional space are shown in Fig. \ref{3dplot}, and it is observed that the desired formation shape is achieved at the specified height starting from an arbitrary initial position. More specifically, Fig. \ref{formation and height} illustrates how the trajectories evolve to achieve the desired formation and reach the specified height of $50$m with gains $k_{pz}=1$ and $k_{vz}=4$.
Fig. \ref{veloity} shows that the UAVs achieve a common final velocity $v_f$ in the horizontal plane, and the velocity corresponding to the $z$ direction decays to zero. The control input required in three directions, to achieve the desired shape, is shown in Fig. \ref{amplitude}. Fig. \ref{phi_theta_psi} depicts the evolution of attitude angles over time.


\begin{figure}
    \centering
    \includegraphics[width=0.8\linewidth]{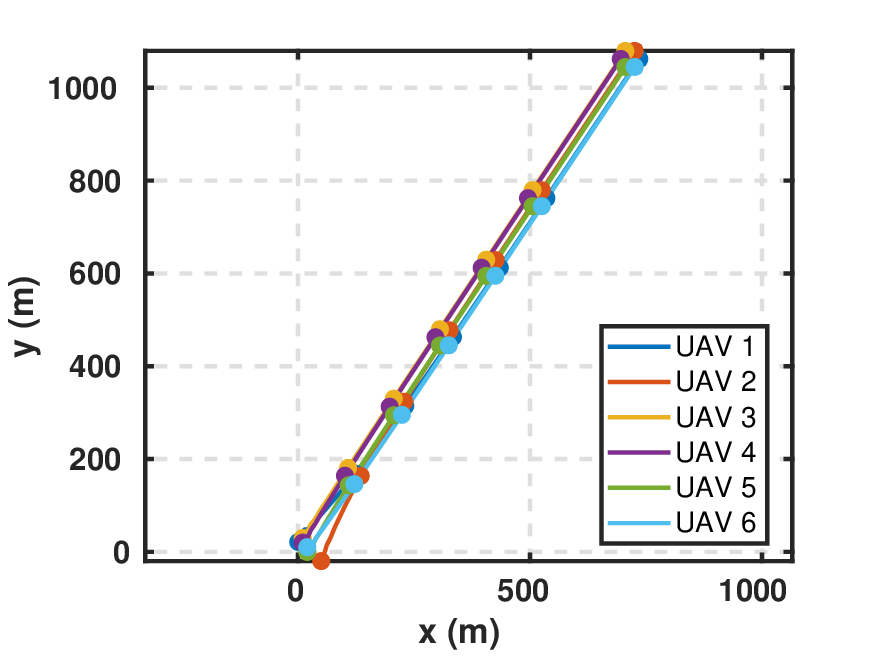}
    \caption{Position of quadrotor UAVs in horizontal plane.}
    \label{xy_2}
\end{figure}

\begin{figure}
    \centering
    \includegraphics[width=0.8\linewidth]{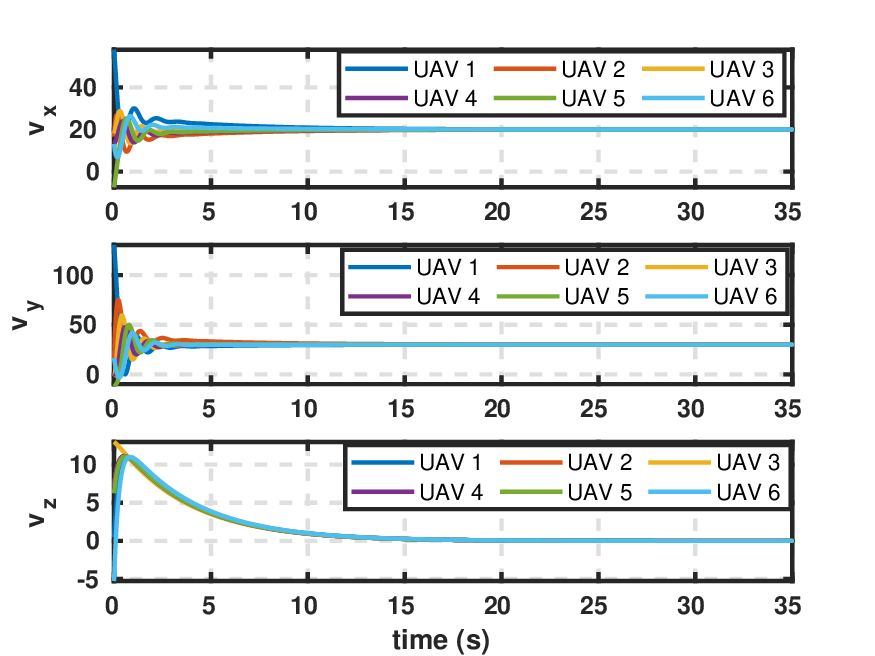}
    \caption{Velocities of quadrotor UAVs.}
    \label{vx_vy_vz_2}
\end{figure}

Suppose it is desired to have a final velocity given by $v_f= \begin{bmatrix}
    20 & 30 
\end{bmatrix}^T$ in the $x$-$y$ plane. The objective is to find a feasible control gain $k$ such that the desired final velocity is achieved starting from the initial velocities mentioned earlier. Equation \eqref{linear_comb} can be represented as
\begin{align*}
    \delta \begin{bmatrix}
        18 \\62
    \end{bmatrix} = \begin{bmatrix}
        -41 \\-97
    \end{bmatrix}.
\end{align*}
No choice of $\delta$ is possible to achieve the desired final velocity with the given initial velocities. Therefore, if a command is given to quadrotor 1 to change its initial velocity to be $v_1(0)=\begin{bmatrix}
    59 & 132.5
\end{bmatrix}^T$, then the desired final velocity is achievable for a choice of $\delta=0.5$. With this choice of $\delta$, the control gain $k$ is evaluated to be $k=-0.5$, a feasible gain as per Theorem \ref{k_theorem}. With this setup, the simulation is performed. It can be seen that the desired formation is achieved, and the quadrotors reach the specified height as shown in Fig. \ref{xy_2}. Furthermore, the desired final flight velocity is achieved as shown in Fig. \ref{vx_vy_vz_2} in the $x$ and $y$ directions as expected. This effectively shows that the final flight velocity can be achieved through a suitably designed control gain $k$.

\section{CONCLUSION}
This work explored the decentralized control of multiple quadrotor vehicles to achieve a desired formation at a specified height. The collective behavior of the group ensured that they fly at some desired speed, while their interaction topology is represented by a weighted ring digraph. Identical macro-vertices of size $2$ were considered in this study, and it was shown that by selecting a single control gain, stability of second-order consensus dynamics over the ring network could be ensured. An analysis of the achievable velocities was carried out, and the results demonstrated that a single control gain aided in determining the common final flight velocity of the overall networked system. Examples were presented to offer deeper insights into the analytical results. Simulations revealed that the desired performance was achieved. In the future, it would be interesting to conduct similar studies for a network with possibly non-identical macro-vertices of size greater than two, over a similar topology.

\bibliography{References}
\begin{IEEEbiography}[{\includegraphics[width=1in,height=1.25in, clip,keepaspectratio]{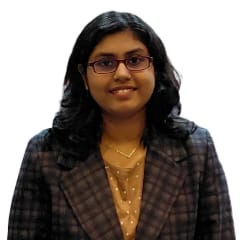}}]{Sahaya Aarti Dennisselvan} received the B.E. degree in Electrical and Electronics Engineering from Government College of Engineering, Tirunelveli, Tamil Nadu, India, in 2014, the M.Tech degree in Instrumentation and Control Systems from National Institute of Technology, Calicut, Kerala, India, in 2021. She is currently working towards her Ph.D. degree
in Electrical Engineering with the Department of
Electrical Engineering, Indian Institute of Technology Bombay, Mumbai, India.
Her research interests include multi-agent systems, networked control, and control theory.
\end{IEEEbiography}
\begin{IEEEbiography}
 [{\includegraphics[width=1in,height=1.25in, clip,keepaspectratio]{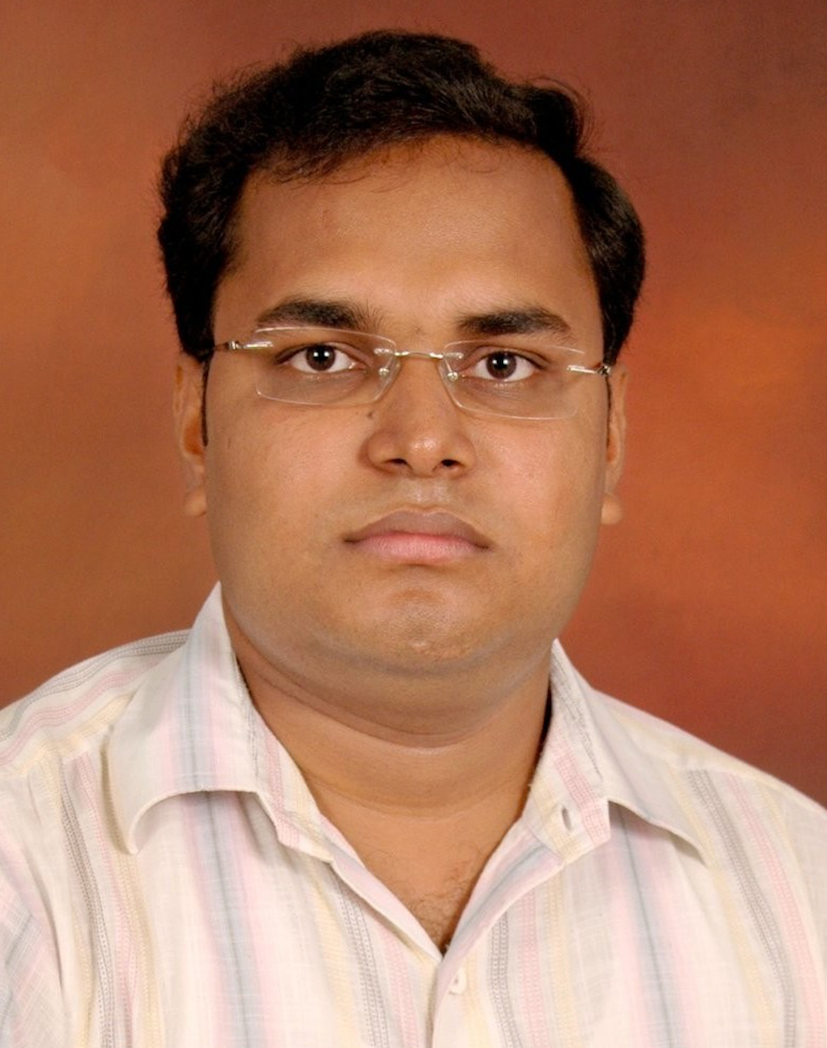}}]{Shashi Ranjan Kumar} received the B.Sc. (Engg.) degree in Electronics and Communication Engineering from the Muzaffarpur Institute of Technology, Muzaffarpur, India, in 2008, and the M.E. and Ph.D. degrees both in Aerospace Engineering from the Indian Institute of Science, Bangalore, India, in 2010 and 2015, respectively. He is an Associate Professor in Aerospace Engineering with the Indian Institute of Technology Bombay, Mumbai, India. From 2015 to 2017, he was a Postdoctoral Fellow with the Faculty of Aerospace Engineering, Technion—Israel Institute of Technology. He is an Associate Editor of IEEE Transactions on Aerospace and Electronic Systems and Proceedings of IMechE Part G: Journal of Aerospace Engineering. His research interests include topics related to guidance of aerospace vehicles, path planning, multi-agent systems, and cooperative control.  
\end{IEEEbiography}
\begin{IEEEbiography}[{\includegraphics[width=1in,height=1.25in,clip,keepaspectratio]{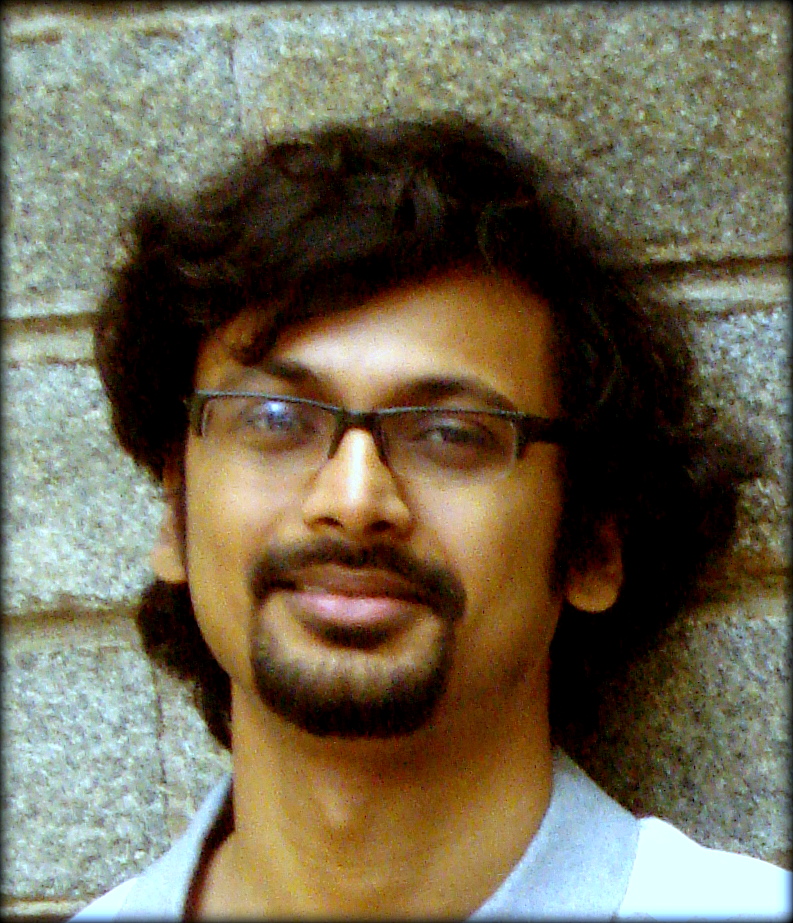}}]{Dwaipayan Mukherjee}
received the B.E. degree in electrical engineering from Jadavpur University, Kolkata, India, in 2007, the M.Tech. degree in control systems engineering from
the Indian Institute of Technology Kharagpur, Kharagpur, India, in 2009, and the Ph.D. degree in engineering from the Department of Aerospace Engineering, Indian Institute of Science, Bengaluru, India, in 2014.
From 2015 to 2017, he was a Postdoctoral Fellow with the Faculty of Aerospace Engineering, Technion—Israel Institute of Technology, Haifa, Israel. He is currently an Associate Professor of Electrical Engineering with the Indian Institute of Technology Bombay, Mumbai, India. His research interests include multi-agent systems, cooperative control, and control theory.
\end{IEEEbiography}

\end{document}